\documentclass[a4paper,11pt]{article}

\pdfoutput=1
\usepackage{jcappub}
\usepackage{aas_macros}

\let\d\relax
\DeclareMathOperator{\d}{d}
\let\L\relax
\DeclareMathOperator{\L}{\mathcal{L}}
\DeclareMathOperator{\order}{\mathcal{O}}
\DeclareMathOperator{\deltaD}{\delta^\mathrm{D}}
\DeclareMathOperator{\WS}{W}
\DeclareMathOperator{\WL}{\mathcal{W}}

\newcommand{\vk}{\mathbf{k}}
\newcommand{\vq}{\mathbf{q}}
\newcommand{\vx}{\mathbf{x}}
\newcommand{\vr}{\mathbf{r}}
\newcommand{\uvk}{{\hat\vk}}
\newcommand{\uvq}{{\hat\vq}}
\newcommand{\uvr}{{\hat\vr}}
\newcommand{\los}{{\hat{\mathbf{n}}}}
\newcommand{\Dist}[1]{D_\mathrm{#1}}
\newcommand{\FAP}{F_\mathrm{AP}}
\newcommand{\nbar}{\bar{n}}
\newcommand{\Plin}{P_\lin}
\newcommand{\Pdimless}{\mathcal{P}}
\newcommand{\D}{\frac{\d\ln\Pdimless}{\d\ln k}}
\newcommand{\Dlin}{\frac{\d\ln\Pdimless_\lin}{\d\ln k}}
\newcommand{\Mpc}{\mathrm{Mpc}}
\newcommand{\Gpc}{\mathrm{Gpc}}
\newcommand{\Mpch}{{\,\Mpc/h}}
\newcommand{\hMpc}{{\,h/\Mpc}}

\renewcommand{\SS}{\mathrm{SS}}
\newcommand{\lin}{\mathrm{lin}}
\newcommand{\eff}{\mathrm{eff}}
\newcommand{\tot}{\mathrm{tot}}
\newcommand{\mock}{\mathrm{mock}}
\newcommand{\sub}{\mathrm{sub}}
\newcommand{\gal}{\mathrm{g}}
\newcommand{\ran}{\mathrm{r}}
\newcommand{\matter}{\mathrm{m}}
\newcommand{\Gauss}{\mathrm{G}}
\newcommand{\nonGauss}{\textsc{ng}}

\title{Galaxy power-spectrum responses and redshift-space super-sample effect}

\author[a,b]{Yin Li,}
\author[c]{Marcel Schmittfull,}
\author[a,d]{\& Uro{\v s} Seljak}

\affiliation[a]{Berkeley Center for Cosmological Physics, Department of
Physics, \& Lawrence Berkeley National Laboratory, University of California,
Berkeley, CA 94720, USA}
\affiliation[b]{Kavli Institute for the Physics and Mathematics of the Universe
(WPI), UTIAS, The University of Tokyo, Chiba 277--8583, Japan}
\affiliation[c]{Institute for Advanced Study, Einstein Drive, Princeton, NJ
08540, USA}
\affiliation[d]{Department of Astronomy, University of California, Berkeley, CA
94720, USA}
\emailAdd{yin.li@berkeley.edu}
\emailAdd{mschmittfull@ias.edu}
\emailAdd{useljak@berkeley.edu}

\abstract{As a major source of cosmological information, galaxy clustering
is susceptible to long-wavelength density and tidal fluctuations.
These long modes modulate the growth and expansion rate of local structures, 
shifting them in both amplitude and scale. These effects are often named the
growth and dilation effects, respectively.
In particular the dilation shifts the baryon acoustic oscillation (BAO) peak
and breaks the assumption of the Alcock-Paczynski (AP) test.
This cannot be removed with reconstruction techniques because the effect
originates from long modes outside the survey.
In redshift space, the long modes generate a large-scale radial peculiar
velocity that affects the redshift-space distortion (RSD) signal.
We compute the \emph{redshift-space} response functions of the
\emph{galaxy} power spectrum to long density
and tidal modes at leading order in perturbation theory, including both the
growth and dilation terms.
We validate these response functions against measurements from
simulated galaxy mock catalogs.
As one application, long density and tidal modes beyond the scale of a survey
correlate various observables leading to an excess error known as the super-sample
covariance, and thus weaken their constraining power.
We quantify the super-sample effect on BAO, AP, and RSD measurements, and
study its impact on current and future surveys.
}

\begin{document}
\maketitle

\section{Introduction}
\label{sec:intro}

Current and upcoming large-scale structure (LSS) surveys offer unprecedented
statistical precision and constraining power on the underlying physics, which
demands equally accurate theoretical predictions.
One of the most important aspects of this is to model the nonlinear coupling
between modes of different wavelengths.
It may exist primordially in the initial conditions, and emerges
gravitationally from structure formation.
In the latter case, the squeezed limit of mode coupling describes the response
of observables of much smaller scales to the long-wavelength perturbations.
A nonzero fluctuation in the mean overdensity of some region modulates the
amplitudes of enclosed smaller structures, and at the same time shifts them
in scales \cite{HamiltonRimesEtAl06, HuKravtsov03, TakadaHu13, LiHuEtAl14}.
These are often named the growth and dilation effects, respectively.

The 2-point function or the power spectrum as its Fourier representation, being
the sole statistics to determine a stationary Gaussian random field, is the
simplest and most commonly used statistics for extracting cosmological
information from LSS probes.
In real space the power spectrum $P(k)$ is a function of only the wavenumber
$k$, due to the statistical homogeneity and isotropy.
A nonzero mean overdensity affects the local $P(k)$ in amplitudes and scales,
so its response includes both growth and dilation terms \cite{LiHuEtAl14,
LiHuEtAl14sss, ChiangWagnerEtAl14, WagnerSchmidtEtAl15a}.
In addition the power spectrum also responds to long-wavelength tidal
perturbations \cite{SchmidtPajerEtAl14, AkitsuTakadaEtAl17, BarreiraSchmidt17},
whose amplitude is of the same order as that of the density mode.
This response depends on the direction of the wavevector $\vk$ as well.
However, given that the Gaussian information is all contained in the estimated
$\hat P(k)$ after averaging over $\uvk$, the mean tidal modes makes no impact
because of the spherical average.

In \emph{redshift-space}, the observed position of \emph{galaxies} is distorted along the
line-of-sight (LOS) by the Doppler effect from their radial peculiar motion
\cite{Kaiser87, Hamilton92}, so the real-space spherical symmetry
is broken to an azimuthal one about the LOS.
Therefore the power spectrum becomes anisotropic and depends additionally on
the angle between $\uvk$ and the LOS, and all the 2-point information is
contained in the azimuthally averaged power spectrum estimator.
As a result the impact of the long-wavelength tides no longer vanishes, and moreover only
one tidal component (out of the 5 degrees of freedom in the traceless symmetric
tensor) contributes due to the azimuthal average.
Both the mean density and tidal fluctuations generate a large-scale
radial peculiar velocity that affects the local RSD
signal\footnote{In addition the radial large-scale bulk flow gives rise to
a systematic offset in the measured redshifts.
However it is suppressed by $\mathcal{H}R$, where $\mathcal{H}^{-1}$ is the
comoving Hubble radius and $R$ is the comoving scale of the bulk flow.
Therefore we ignore this effect here.
} \cite{FeixNusser13}.
The full response of the \emph{redshift-space galaxy} 2-point function combines the
effects of density, tide, and peculiar velocity, as well as galaxy biasing.
We expect it to again consist of growth and dilation pieces.

For a LSS survey, a super-survey density fluctuation coherently changes the
real-space matter $n$-point functions according to their responses, therefore
introducing additional noise due to its stochasticity, known as the super-sample
covariance (SSC) \cite{HamiltonRimesEtAl06, SefusattiCrocceEtAl06,
TakadaBridle07, TakadaJain09, SatoEtAl09, TakahashiEtAl09, SchneiderColeEtAl11,
dePutterEtAl12, TakadaHu13, LiHuEtAl14, MohammedSeljak14, SchaanEtAl14,
MohammedSeljakVlah17, HowlettPercival17, ChanDizgahEtAl17}.
Alternatively it can be modeled as an extra parameter that is degenerate with
other cosmological parameters and degrades their constraints \cite{LiHuEtAl14sss}.
The two views are equivalent.
The latter is easier to implement in the analysis \cite{LiHuEtAl14sss}, even
though the SSC treatment is more well-received in the literature.
Similarly, in \emph{redshift space} both the super-survey density and tidal modes will
generate super-sample covariance of the \emph{galaxy} power spectrum, which will again
degrade cosmological constraints from galaxy clustering.
The full response functions, including both growth and dilation effects, will be
relevant to the RSD measurement of the growth rate.
And the dilation effect alone will stretch all scales isotropically (by the mean
density) or anisotropically (by the mean tide), so it could be important to geometric
probes like the BAO peak and AP effect.
This cannot be removed with reconstruction techniques \cite{EisensteinSeoEtAl07}
as the effect originates from long modes beyond the survey scale.
As an example of the dilation effect, it has been shown that by considering a local underdense
environment one can partially relieve the tension between the measured local
expansion rate $H_0$ and its value inferred from the Cosmic Microwave
Background \cite{MarraAmendolaEtAl13, WojtakKnebeEtAl14, Ben-DayanDurrerEtAl14,
IchikiYooEtAl16, OdderskovKoksbangEtAl16}.
More generally for multiple surveys, their super-survey modes are correlated
depending on their geometries and relative locations, leading to
super-sample covariance even between surveys that are not physically
overlapping.

The previous literature have studied extensively the effect of the long modes
on the power spectrum in real space.
In this paper we present for the first time this effect in redshift space,
which is more realstic observationally and therefore allow us to quantify its
impact on galaxy surveys.
The rest of the paper is organized as follows.
In Sec.~\ref{sec:ana}, we derive the power spectrum response functions to mean
density and tidal modes in real space (for matter) and redshift space (for
galaxies) with tree-level perturbation theory, and demonstrate the shift of
scales due to the redshift-space dilation effect.
We numerically measure the response functions from simulated galaxy mock
catalogs and compare
them to our analytical results in Sec.~\ref{sec:mock}.
Sec.~\ref{sec:ss} discusses the super-sample effect as a result of the
responses, and its impact on constraints by galaxy clustering measurements.
We conclude in Sec.~\ref{sec:con}.

\subsection{Notation}
\label{sub:notation}

In this paper we use the following shorthand notation for configuration-space (over
variable $\vx$ or $\vr$) and Fourier-space (over variable $\vk$ or $\vq$)
integrals
\begin{equation*}
    \int_\vx \longrightarrow \int\!\d^3\vx,
    \qquad
    \int_\vk \longrightarrow \int\!\frac{\d^3\vk}{(2\pi)^3}.
\end{equation*}
We adopt the notation
\begin{equation*}
    \int_{\vk\in V_{k_i}} \longrightarrow \int\frac{\d^3\vk}{V_{k_i}},
\end{equation*}
for binning over the $i$th spherical $\vk$-shell with width $\Delta
k_i=k_{i+1}-k_i$ and volume $V_{k_i}=\frac{4\pi}3(k_{i+1}^3-k_i^3)$.
In the zero-width limit it reduces to the average over solid angle, that we
abbreviate by
\begin{equation*}
    \int_\uvk \longrightarrow \int\!\frac{\d\Omega_\vk}{4\pi}.
\end{equation*}

To distinguish quantities that have conventional name collisions, we denote the
power spectrum by $P$, while using $\L_l$ as the Legendre polynomial of order
$l$.
And $\Pdimless\equiv k^3 P/2\pi^2$ is the dimensionless power spectrum, whereas
$\Delta_L$ is the amplitude of the mean density ($L=0$) or tidal ($L=2$)
fluctuation.
For conciseness, we drop the subscript of power spectrum that labels matter or
galaxy field, and let $P(k)$ denote the \emph{real-space matter} power
spectrum, while $P(k,\mu)$ or $P_l(k)$ stands for the \emph{redshift-space
galaxy} power spectrum.

For a LSS survey with window function $\WS(\vx)$, we use the following
notation for various volume normalization factors
\begin{equation}
    \label{Vn}
    V_n \equiv \int_\vx \WS^n(\vx).
\end{equation}
Therefore
\begin{equation}
    \label{V2}
    V_2 \equiv \int_\vx \WS^2(\vx) = \int_\vq |\WS(\vq)|^2.
\end{equation}
We denote the rough size of the survey by $V^\frac13$.

\section{Responses by perturbative calculation}
\label{sec:ana}

First we define concretely the long-wavelength and short-wavelength modes using
a window function in Sec.~\ref{sub:modes}.
Then we compute the response functions of the real-space matter power spectrum
in Sec.~\ref{sub:resp_real}, and those of the redshift-space galaxy power
spectrum in Sec.~\ref{sub:resp_red}, at tree-level with standard perturbation
theory.
In Sec.~\ref{sub:dilation} we derive the shift of scales generated by the
dilation effect of the long density and tidal modes.

\subsection{Long and short modes}
\label{sub:modes}

There are two common usages of a window function, either to mask a field to
limit oneself to observables below the window scale, or to smooth it to focus
on larger structures.
The masking and smoothing scales coincide when we quantify the long and short
modes.
The same window serves to describe both the geometry of the observed
short-wavelength sample, and the scales beyond which long-wavelength
fluctuation arises.

For the matter overdensity field $\delta(\vx)=\rho(\vx)/\bar\rho-1$ where
$\rho(\vx)$ is the matter density and $\bar\rho$ is its mean value, the masked
field with a superscript $\WS$ is a product in configuration space
\begin{equation}
    \label{short}
    \delta^W(\vx) \equiv \delta(\vx) \WS(\vx),
\end{equation}
where $\WS$ is the masking window.
A similar relation holds for the galaxy field, with $\delta(\vx)$ replaced by
$\delta_\gal(\vx)$.
As usual $\WS(\vx)\propto \nbar(\vx)w(\vx)$ where $\nbar$ is the selection
function describing the sample geometry, and $w$ is the statistical weight,
e.g.\ the FKP weight \cite{FeldmanKaiserEtAl94} assigned to each galaxy.

On the other hand, the smoothed long mode is a convolution of some field with
the window function in configuration space, therefore a product in Fourier
space.
We find the following unified definition of long modes works for both the mean
density ($L=0$) and tidal fluctuations ($L=2$)
\begin{equation}
    \label{long}
    \Delta_L(\uvk) \equiv \frac{1}{V_2} \int_\vq
    \L_L(\uvk\cdot\uvq) \delta(\vq)\WL(-\vq),
\end{equation}
where $\WL$ is the smoothing window, $\L_L$ is Legendre polynomial of order
$L$, and $V_2$ is a normalization volume defined in \eqref{Vn}.
We will see later that for power spectrum responses, the smoothing window is
related to the masking one by $\WL(\vx)=\WS^2(\vx)$ for dimensionless window
functions, which holds trivially in cases of uniform windows as in
Ref.~\cite{TakadaHu13, ChiangWagnerEtAl14}.

Obviously $\Delta_0$ is the mean overdensity within the window.
For $L=2$, \eqref{long} gives the mean tide projected to the direction of
$\uvk$, multiplied by a factor of $3/2$ from $\L_2$.
Note that the tidal tensor has 5 degrees of freedom, for which we can pick the
spherical harmonic basis
\begin{equation}
    \Delta_{LM} \equiv \frac{\sqrt{4\pi}}{V_2}
    \int_\vq {Y_L^M}^*(\uvq) \delta(\vq)\WL(-\vq),
\end{equation}
to decompose it into
\begin{equation}
    \label{tide_decomp}
    \Delta_2(\uvk) = \frac{\sqrt{4\pi}}5 \sum_{M=-2}^2 Y_2^M(\uvk) \Delta_{2M}.
\end{equation}
As we will see later, given the azimuthal symmetry the only relevant tidal mode
is the combination in \eqref{tide_decomp}.
This is the case for $P(\vk)$ response in real space for $kV^\frac13\gg1$, where
the direction of $\uvk$ breaks the spherical symmetry to the azimuthal symmetry
before angular average.
Also in redshift space, the power spectrum averaged azimuthally about the
line-of-sight (LOS) direction $\los$ will only respond to the projected mode
$\Delta_2(\los)$ but not directly to each $\Delta_{2M}$.

\subsection{Real-space matter responses}
\label{sub:resp_real}

As an example with simpler physics, let's first review the effects of long
modes on the matter 2-point function in the real space.
As usual the power spectrum is defined by
\begin{equation}
    \bigl\langle \delta(\vk)\delta(\vk') \bigr\rangle
    \equiv (2\pi)^3 \deltaD(\vk+\vk') P(k)
\end{equation}
where $\langle\dots\rangle$ denotes the ensemble average over all possible
realizations and the Dirac delta function $\deltaD$ results from the
translational invariance, which is broken in the presence of a window.
The masked short modes in \eqref{short} become a convolution in Fourier space
\begin{equation}
    \label{short_k}
    \delta^W(\vk) = \int_\vq \delta(\vk+\vq) \WS(-\vq),
\end{equation}
where $\WS(\vq)$ is the masking window in Fourier space, and is suppressed on
scales beyond $qV^\frac13\lesssim1$.
It's straightforward to show that the variance of a masked short mode is the
convolved power spectrum
\begin{equation}
    \label{P}
    \bigl\langle \delta^W(\vk)\delta^W(-\vk) \bigr\rangle
    = \int_\vq P(|\vk-\vq|) |\WS(\vq)|^2 \simeq V_2 P(k).
\end{equation}
And the approximation holds in the limit $kV^{\frac13}\gg1$ where the window
effect becomes isotropic on scales much smaller than its size.
As a result of this spherical symmetry, the power spectrum on the right hand
side is only a function of scale.

Now let's define a local 2-point function that samples all realizations of the
short modes for a single realization of the long ones.
To the leading order, the local 2-point function will receive corrections
proportional to $\Delta_0$ and $\Delta_2$
\begin{equation}
    \label{mod}
    \bigl\langle \delta^W(\vk)\delta^W(-\vk) \bigr\rangle_W
    \simeq
    \bigl\langle \delta^W(\vk)\delta^W(-\vk) \bigr\rangle
    \bigl[ 1 + \order(\Delta_L) \bigr].
\end{equation}
Here we distinguish a local ensemble average $\langle\dots\rangle_W$ where
long modes are held fixed from the usual global ensemble average.
Compared to \eqref{P}, the density modulation $\order(\Delta_0)$ is still
isotropic, while the tidal modulation $\order(\Delta_2)$ depends on $\uvk$ even
though the global 2-point function is isotropic.
Our goal is to identify the response function $R_L(k)$ from the fractional
modulation $\order(\Delta_L) \sim R_L\Delta_L$.
We can write this real-space response function as
\begin{equation}
    \label{resp_real}
    R_L(k) \equiv \frac{\d\ln P^W(\vk)}{\d\Delta_L(\uvk)},
    \quad L=0,2.
\end{equation}
Note that due to symmetry the leading order tidal modulation in $P^W(\vk)$
should have the same $\uvk$ dependence as that in $\Delta_2(\uvk)$, so that
$R_2(k)$ is only a function of scale.

\eqref{mod} implies that one can compute the response functions by correlating
the 2-point function of the short modes with one long mode
$\langle\delta^W(\vk)\delta^W(-\vk)\Delta_L\rangle$, which is a squeezed
3-point correlation.
Alternatively we have also obtained the same result from the quadrilaterally
squeezed 4-point function $\langle\delta^W(\vk)\delta^W(-\vk)
\delta^W(\vk')\delta^W(-\vk')\rangle$.
In both approaches we read off the response functions from terms proportional
to the variance of the long modes, given later in \eqref{long_cov}.
For simplicity we only show the 3-point calculations below.

Substituting the expression for the long and short modes in \eqref{long} and
\eqref{short_k} into the squeezed 3-point correlation
\begin{multline}
    \label{3pt_real}
    \big\langle \delta^W(\vk)\delta^W(-\vk)\Delta_L(\uvk) \big\rangle
    = \frac{1}{V_2} \int_{\vq\vq'} B(\vq, \vk+\vq', -\vk-\vq-\vq') \\
    \times \L_L(\uvk\cdot\uvq) \WL(-\vq)\WS(\vq+\vq')\WS(-\vq'),
\end{multline}
in which the bispectrum is defined by
\begin{equation}
    \label{bispec}
    \bigl\langle \delta(\vk_1)\delta(\vk_2)\delta(\vk_3) \bigr\rangle
    \equiv (2\pi)^3 \deltaD(\vk_1+\vk_2+\vk_3) B(\vk_1, \vk_2, \vk_3).
\end{equation}

At leading order, the tree-level bispectrum is determined by the linear power spectrum $\Plin$ and the $F_2$ kernel, which has a specific dependence on the angle between wavevectors that involves Legendre polynomials of order $L=0$, 1 and 2.
Because the window scale separates the wavelength of long and short modes
$q\lesssim V^{-\frac13}\lesssim k$, we can carry out the calculation in the
squeezed limit $q/k\ll1$.
See more technical details of the calculation in App.~\ref{sec:spt}.
Putting all the pieces from \eqref{Btree_real}, \eqref{F2_exp}, and
\eqref{Plin_exp} together,
\begin{align}
    \label{Btree_real1}
    & B(\vq, \vk+\vq', -\vk-\vq-\vq') \nonumber\\
    &\simeq
    2 F_2(\vk+\vq', \vq) \Plin(\vk+\vq')\Plin(q) +
    2 F_2(-\vk-\vq'-\vq, \vq) \Plin(-\vk-\vq'-\vq)\Plin(q) \nonumber\\
    &\simeq \Plin(k) \Plin(q)
    \biggl[ \Bigl( \frac{68}{21} - \frac13 \Dlin \Bigr) +
    \L_2(\uvk\cdot\uvq) \Bigl( \frac{58}{21} - \frac23 \Dlin \Bigr) \biggr],
\end{align}
which involves only $L=0$ and $L=2$ Legendre polynomials.
To calculate \eqref{3pt_real}, we can integrate out $\vq'$ using the convolution theorem,
\begin{equation}
    \int_{\vq'} \WS(\vq+\vq') \WS(-\vq') = \int_\vx \WS^2(\vx) e^{-i\vq\cdot\vx}.
\end{equation}
If we identify $\WS^2(\vx)$ with $\WL(\vx)$, \eqref{3pt_real}
then simplifies to
\begin{multline}
    \label{3pt_real1}
    \big\langle \delta^W(\vk)\delta^W(-\vk)\Delta_L(\uvk) \big\rangle
    \simeq V_2 \Plin(k) \\
    \times
    \biggl[ \sigma_{L0}(\uvk) \Bigl( \frac{68}{21} - \frac13 \Dlin \Bigr) +
    \sigma_{L2}(\uvk) \Bigl( \frac{58}{21} - \frac23 \Dlin \Bigr) \biggr],
\end{multline}
where the relation $\WS^2(\vx)=\WL(\vx)$ has helped to complete the expected
form of the long mode covariance
\begin{equation}
    \label{long_cov}
    \sigma_{LL'}(\uvk)
    \equiv \bigl\langle\Delta_L(\uvk)\Delta_{L'}(\uvk)\bigr\rangle
    = \frac{1}{V_2^2} \int_\vq
    \L_L(\uvk\cdot\uvq)\L_{L'}(\uvk\cdot\uvq)
    \Plin(q) \bigl|\WL(\vq)\bigr|^2.
\end{equation}
Physically the relation $\WS^2(\vx)=\WL(\vx)$ results from the fact that we are
considering the long modes that modulate a 2-point function, and therefore they
need to be weighted by the masking window squared.
Also note that one of the $\L_L$ factors in \eqref{3pt_real1} and
\eqref{long_cov} originates from \eqref{long}, and constitutes the right
angular weight in the long mode covariance.
This justifies the definition of the long modes, and our previous
symmetry-based argument that only the projected tidal mode matters.
Obviously, $\sigma_{00}$ is the variance of mean overdensity.
For an isotropic window, the long mode covariance matrix $\sigma_{LL'}$ is diagonal, isotropic,
and is simply related to $\sigma_{00}$ by
\begin{equation}
    \label{long_cov_iso}
    \sigma_{LL'} = \frac{\delta_{LL'}\sigma_{00}}{2L+1}.
\end{equation}
From \eqref{3pt_real1} one can easily read off the response functions
\begin{align}
    \label{R0R2}
    R_0(k) &= \frac{68}{21} - \frac13 \Dlin, \nonumber\\
    R_2(k) &= \frac{58}{21} - \frac23 \Dlin.
\end{align}

The extensively studied density response $R_0(k)$ has a growth term and a
dilation term.
The former arises from a growth modulation by the mean density -- more
structure forms in denser environment, and it is a constant $68/21$ on large
scale.
The latter piece is due to a change of local expansion history -- all
scales shifts isotropically depending on the mean overdensity.
We will demonstrate the dilation scale-shift explicitly in Sec.~\ref{sub:dilation}.

Similarly, the tidal response $R_2(k)$ is also composed of the growth and
dilation effects.
$\Delta_2$ generates an ellipsoidal deviation from the isotropic background
expansion.
This introduces a direction-dependent shift in scales leading to the dilation
term.
And the anisotropic local expansion further induces a direction-dependent
modulation in the local linear growth function giving rise to a constant growth
term $58/21$ on large scale.
Our result is consistent with previous derivations \cite{SchmidtPajerEtAl14,
AkitsuTakadaEtAl17} on $R_2(k)$ except for a factor of $2/3$ due to the
difference in definition \eqref{long}.

\subsection{Redshift-space galaxy responses}
\label{sub:resp_red}

Assuming the global plane-parallel approximation where the LOS $\los$ is fixed
across the survey region, the estimator of the redshift-space galaxy power
spectrum is simply
\begin{equation}
    \label{P_mu}
    \hat P^W(k,\mu) \equiv \frac{2}{V_2}
    \int_\uvk \delta_\gal^W(\vk) \delta_\gal^W(-\vk) \deltaD(\mu-\uvk\cdot\los).
\end{equation}
This can be represented alternatively by its multipole moments
\begin{equation}
    \label{P_l}
    \hat P_l^W(k) \equiv \frac{2l+1}{V_2}
    \int_\uvk \delta_\gal^W(\vk) \delta_\gal^W(-\vk) \L_l(\uvk\cdot\los).
\end{equation}
The two representations are related by the multipole expansion of the Dirac delta
\begin{equation}
    \label{deltaD_l}
    \deltaD(\mu-\uvk\cdot\los) = \sum_{l=0}^\infty
    \frac{2l+1}{2} \L_l(\mu) \L_l(\uvk\cdot\los).
\end{equation}
The subscript g denotes the galaxy field in redshift space.
In practice, the $\uvk$-integral should in addition span a shell of finite
width around $k$, and the Dirac delta in $\mu$ should also be replaced by
finite binning.

In real space the ensemble average of \eqref{P_l} only has a monopole ($l=0$)
and is only modulated by $\Delta_0$ due to isotropy.
RSD breaks the spherical symmetry to an azimuthal one about $\los$, introducing
additional angular dependence on $\mu$ (or $l$).
While now the long tidal modes also matters, the azimuthal symmetry determines
that only their projected component along $\los$, i.e.\ $\Delta_2(\los)$, is
relevant out of 5 degrees of freedom, given $\los$ being the symmetry breaking
direction.

Compared to the real-space calculation in Sec.~\ref{sub:resp_real}, in redshift
space we also need to model the galaxy biasing and the redshift-space
distortion.
For the biasing we relate the real-space galaxy field to the underlying matter
overdensity with the linear bias coefficient $b_1$ \cite{Kaiser84}, the second
order local bias $b_2$ \cite{FryGaztanaga93}, and the non-local tidal bias
$b_{s^2}$ \cite{McDonaldRoy09, BaldaufSeljakEtAl12, ChanScoccimarroEtAl12}
\begin{equation}
    \label{bias}
    \delta_\gal(\vx) = b_1 \delta(\vx) + \frac{b_2}2 \delta^2(\vx)
    + \frac{b_{s^2}}2 s_{ij}(\vx)s_{ij}(\vx),
\end{equation}
where the tidal field $s_{ij}(\vx) \equiv \partial^2
\phi(\vx)/\partial x_i\partial x_j - \nabla^2\phi/3$ is the traceless Hessian
of the gravitational potential $\phi(\vx)$ determined by the Poisson equation
$\nabla^2\phi(\vx) = 4\pi \delta(\vx)$.
We then model the linear RSD effect \cite{Kaiser87, Hamilton92} for the
tree-level galaxy bispectrum \cite{ScoccimarroCouchmanEtAl99}.
Combining everything is equivalent to using the $Z_1$ and $Z_2$ kernels instead
of $F_2$.
For more details see App.~\ref{sec:spt}.

The redshift-space response functions $R_L(k,\mu)$ capture the modulation of the local power
spectrum by the long modes,
\begin{equation}
    \label{localP_mu}
    \langle\hat P^W(k,\mu)\rangle_W \simeq P(k,\mu)
    \bigl[1+{\textstyle\sum_L}R_L(k,\mu)\Delta_L\bigr],
\end{equation}
around the global average $P(k,\mu)$, which for linear RSD reads
\begin{equation}
    P(k,\mu) = (b_1+f\mu^2)^2 \Plin(k)
\end{equation}
with only 3 nonzero multipole moments,
\begin{equation}
    \label{Kaiser}
    \begin{bmatrix}
        P_0(k) \\ P_2(k) \\ P_4(k)
    \end{bmatrix} = \Plin(k)
    \begin{bmatrix}
        b_1^2 + 2b_1f/3 + f^2/5 \\ 4b_1f/3 + 4f^2/7 \\ 8f^2/35
    \end{bmatrix}.
\end{equation}
The growth rate $f\equiv\d\ln D/\d\ln a$ is the derivative of the linear growth
function $D(a)$.
We should emphasize that $\Delta_L$ are still the long modes of the matter field in
real space given they are not directly measurable.

Just like the real-space response functions, the redshift-space responses
should also consist of modulations in amplitudes and scales, corresponding again to
 growth and dilation pieces.
Similar to \eqref{resp_real} we define
\begin{equation}
    \label{Rmu}
    R_L(k,\mu) \equiv \frac{\d\ln P^W(k,\mu)}{\d\Delta_L(\los)}
    = G_L + D_L \D,
\end{equation}
where $G_L$ and $D_L$ are growth and dilation coefficients,
which in general can depend on both $k$ and $\mu$.
At leading order they are only functions of $\mu$ as we will derive below.
Alternatively, we can also define the response of power spectrum multipoles
\begin{equation}
    \label{Rl}
    R_{lL}(k) \equiv \frac{\d\ln P^W_l(k)}{\d\Delta_L(\los)}
    = G_{lL} + D_{lL} \D,
\end{equation}
where now the growth and dilation coefficients $G_{lL}$ and $D_{lL}$ are
constants at leading order.
The two representations can be related by
\begin{equation}
    \label{R_trans}
    R_{lL}(k) P_l(k) = \frac{2l+1}{2} \int_{-1}^1\!\!\d\mu\;
    R_L(k,\mu) P(k,\mu) \L_l(\mu).
\end{equation}

Now let's consider the following squeezed 3-point correlation, and plug in the
definitions \eqref{long}, \eqref{short_k}, and \eqref{P_mu}
\begin{multline}
    \label{3pt_red}
    \big\langle \hat P^W(k,\mu) \Delta_L(\los) \big\rangle
    = \frac{2}{V_2^2} \int_{\uvk\vq\vq'} B_{\matter\gal\gal}(\vq, \vk+\vq', -\vk-\vq-\vq') \\
    \times \deltaD(\mu-\uvk\cdot\los) \L_L(\uvq\cdot\los) \WL(-\vq)\WS(\vq+\vq')\WS(-\vq'),
\end{multline}
Using \eqref{Btree_red}, \eqref{Z2_exp} and \eqref{Z1Plin_exp}, we expand
the tree-level redshift-space galaxy bispectrum
in the squeezed limit ($q/k\ll 1$)
 as follows:
\begin{align}
    \label{Btree_red1}
    & B_{\matter\gal\gal}(\vq, \vk+\vq', -\vk-\vq-\vq')
    \simeq 2 Z_2(\vk+\vq', \vq) Z_1(\vk+\vq') \Plin(\vk+\vq')\Plin(q) \nonumber\\
    &\hspace{20ex}
    + 2 Z_2(-\vk-\vq'-\vq, \vq) Z_1(-\vk-\vq'-\vq) \Plin(-\vk-\vq'-\vq)\Plin(q) \nonumber\\
    &\simeq (b_1 + f\mu^2) \Plin(k) \Plin(q) \biggl\{ \nonumber\\
        &\qquad \Bigl( \frac{68}{21}b_1 + \frac13b_1f + 2b_2
            + 2b_1f\mu^2 + \frac{52}{21}f\mu^2 - \frac13f^2\mu^2
            - \frac13 (b_1 + f\mu^2) \Dlin \Bigr) \nonumber\\
        &\qquad + \Bigl( \frac{58}{21}b_1 + \frac43b_{s^2} + \frac{74}{21}f\mu^2
            - \frac23 (b_1 + f\mu^2) \Dlin \Bigr) \L_2(\uvk\cdot\uvq)
        + \frac23 f (b_1 - f\mu^2) \L_2(\uvq\cdot\los) \nonumber\\
        &\qquad + f\mu \Bigl( 3b_1 + 7f\mu^2 - (b_1 + f\mu^2) \Dlin \Bigr)
            (\uvq\cdot\los) (\uvk\cdot\uvq)
    \biggr\}.
\end{align}

Plugging it back into \eqref{3pt_red}, we can immediately integrate out $\uvk$,
$\vq$, and $\vq'$ with the help of relation\footnote{This follows
straightforwardly from \eqref{deltaD_l}, the addition theorem and the
orthonormality of spherical harmonics.}
\begin{equation}
    2\int_\uvk \L_l(\uvk\cdot\uvq) \deltaD(\mu-\uvk\cdot\los)
    = \L_l(\mu) \L_l(\uvq\cdot\los),
\end{equation}
to derive
\begin{align}
    \label{3pt_red1}
    & \big\langle\hat P^W(k,\mu)\Delta_L(\los)\big\rangle
    = (b_1 + f\mu^2)\Plin(k) \biggl\{ \nonumber\\
        &\hspace{1.5em} \Bigl(
            \frac{68}{21}b_1 + \frac13b_1f + 2b_2
            + 3b_1f\mu^2 + \frac{52}{21}f\mu^2 -\frac13f^2\mu^2
            + \frac73f^2\mu^4 \nonumber\\
            &\hspace{3em} - \frac13 (1 + f\mu^2) (b_1 + f\mu^2) \Dlin
        \Bigr) \sigma_{L0}(\los) \nonumber\\
        &\hspace{0.4em} + \Bigl(
            - \frac{29}{21}b_1 + \frac23b_1f - \frac23b_{s^2}
            + \frac{29}7b_1\mu^2 + 2b_1f\mu^2 - \frac{37}{21}f\mu^2 - \frac23f^2\mu^2 + 2b_{s^2}\mu^2
            + \frac{37}7f\mu^4 + \frac{14}3f^2\mu^4 \nonumber\\
            &\hspace{3em} - \frac13 (3\mu^2 + 2f\mu^2 - 1) (b_1 + f\mu^2) \Dlin
        \Bigr) \sigma_{L2}(\los)
    \biggr\},
\end{align}
from which we can read off $R_L(k,\mu)$ easily by identifying pieces
proportional to $\sigma_{LL}(\los)$.
In terms of growth and dilation coefficients, the responses to the mean density are
\begin{align}
    \label{GD_mu}
    G_0(\mu) &= \frac{\frac{68}{21}b_1 + \frac13b_1f + 2b_2
        + \bigl( 3b_1f + \frac{52}{21}f -\frac13f^2 \bigr) \mu^2 + \frac73f^2\mu^4}
        {b_1 + f\mu^2}, \nonumber\\
    D_0(\mu) &= -\frac13 (1 + f\mu^2),
\end{align}
while the responses to the tide are
\begin{align}
    \label{GD_mu2}
    G_2(\mu) &= \frac{- \frac{29}{21}b_1 + \frac23b_1f - \frac23b_{s^2}
        + \bigl( \frac{29}7b_1 + 2b_1f - \frac{37}{21}f - \frac23f^2 + 2b_{s^2} \bigr) \mu^2
        + \bigl( \frac{37}7f + \frac{14}3f^2 \bigr) \mu^4}{b_1 + f\mu^2}, \nonumber\\
    D_2(\mu) &= -\frac13 (3\mu^2 + 2f\mu^2 - 1).
\end{align}

\begin{table}[tbp]
    \centering
    \caption{Growth coefficients of responses of the redshift-space galaxy
    power spectrum multipoles $P_l(k)$ to the mean density ($L=0$), and mean
    tide ($L=2$).}
    \label{tab:G_l}
    \begin{tabular}{c|c}
        \hline
        & $G_{lL}$ \\
        \hline
        $l=0,L=0$
        & $\displaystyle\frac{\frac{68}{21}b_1^2 + \frac43b_1^2f + \frac{40}{21}b_1f
            + \frac{16}{15}b_1f^2 + \frac{52}{105}f^2 + \frac4{15}f^3 + 2b_1b_2 + \frac23b_2f}
        {b_1^2 + \frac23b_1f + \frac15f^2}$ \\[1em]
        $l=2,L=0$
        & $\displaystyle\frac{2b_1^2f+\frac{80}{21}b_1f + \frac{64}{21}b_1f^2
            + \frac{208}{147}f^2 + \frac{58}{63}f^3 + \frac43b_2f}
        {\frac43b_1f + \frac47f^2}$ \\[1em]
        $l=4,L=0$
        & $\displaystyle\frac{\frac{128}{105}b_1f^2 + \frac{416}{735}f^2 + \frac{752}{1155}f^3}
        {\frac8{35}f^2}$ \\[1em]
        $l=0,L=2$
        & $\displaystyle\frac{\frac43b_1^2f + \frac{88}{105}b_1f + \frac43b_1f^2
            + \frac{296}{735}f^2 + \frac8{15}f^3 + \frac8{45}fb_{s^2}}
        {b_1^2 + \frac23b_1f + \frac15f^2}$ \\[1em]
        $l=2,L=2$
        & $\displaystyle\frac{\frac{58}{21}b_1^2 + \frac43b_1^2f + \frac{484}{147}b_1f
            + \frac{80}{21}b_1f^2 + \frac{74}{49}f^2 + \frac{116}{63}f^3
            + \frac43b_1b_{s^2} + \frac{44}{63}fb_{s^2}}
        {\frac43b_1f + \frac47f^2}$ \\[1em]
        $l=4,L=2$
        & $\displaystyle\frac{\frac{528}{245}b_1f + \frac{32}{21}b_1f^2 + \frac{10064}{8085}f^2
            + \frac{1504}{1155}f^3 + \frac{16}{35}fb_{s^2}}
        {\frac8{35}f^2}$ \\
        \hline
    \end{tabular}
\end{table}

\begin{table}[tbp]
    \centering
    \caption{Dilation coefficients of responses of the redshift-space galaxy
    power spectrum multipoles $P_l(k)$ to the mean density ($L=0$), and mean
    tide ($L=2$).}
    \label{tab:D_l}
    \begin{tabular}{c|c}
        \hline
        & $D_{lL}$ \\
        \hline
        $l=0,L=0$
        & $\displaystyle-\frac{\frac13b_1^2 + \frac19b_1^2f + \frac29b_1f
            + \frac2{15}b_1f^2 + \frac1{15}f^2 + \frac1{21}f^3}
        {b_1^2 + \frac23b_1f + \frac15f^2}$ \\[1em]
        $l=2,L=0$
        & $\displaystyle-\frac{\frac29b_1^2f + \frac49b_1f + \frac8{21}b_1f^2
            + \frac4{21}f^2 + \frac{10}{63}f^3}
        {\frac43b_1f + \frac47f^2}$ \\[1em]
        $l=4,L=0$
        & $\displaystyle-\frac{\frac{16}{105}b_1f^2 + \frac8{105}f^2 + \frac8{77}f^3}
        {\frac8{35}f^2}$ \\[1em]
        $l=0,L=2$
        & $\displaystyle-\frac{\frac29b_1^2f + \frac8{45}b_1f + \frac4{15}b_1f^2
            + \frac8{105}f^2 + \frac2{21}f^3}
        {b_1^2 + \frac23b_1f + \frac15f^2}$ \\[1em]
        $l=2,L=2$
        & $\displaystyle-\frac{\frac23b_1^2 + \frac49b_1^2f + \frac{44}{63}b_1f
            + \frac{16}{21}b_1f^2 + \frac27f^2 + \frac{20}{63}f^3}
        {\frac43b_1f + \frac47f^2}$ \\[1em]
        $l=4,L=2$
        & $\displaystyle-\frac{\frac{16}{35}b_1f + \frac{32}{105}b_1f^2 + \frac{272}{1155}f^2
            + \frac{16}{77}f^3}
        {\frac8{35}f^2}$ \\
        \hline
    \end{tabular}
\end{table}

We can also transform the responses into their multipole representation via
\eqref{R_trans}, and tabulate the results in Tabs.~\ref{tab:G_l} and
\ref{tab:D_l}.
Note that we have not factored out the $f$'s in the fractions in these tables, so that they
serve as reminders that those responses are ill-defined when $f\to0$ and do not
reduce to the correct real-space limit.
In this limit the $l\neq0$ denominators vanish and one should instead normalize
with $b_1^2$ (from $P_0$), and then can easily verify that the redshift-space
reponses return to the real-space results \eqref{R0R2}, with $R_{00}\to R_0$
and $R_{22}\to R_2$ respectively, for unbiased ($b_1=1$, $b_2=b_{s^2}=0$)
tracers.

Notice that all numerators are proportional to $f$ for $l\neq L$ suggesting
these responses are purely a RSD effect.
This is because in real space a long-wavelength monopole fluctuation ($L=0$)
can only give rise to a monopole ($l=0$) and likewise for quadrupole, as we
have already shown in Sec.~\ref{sub:resp_real}.

In addition to the tabulated $l=0,2,4$ cases there is a remaining correlation in
\eqref{3pt_red1} for $l=6$,
\begin{multline}
    \big\langle\hat P^W_6(k)\Delta_L(\los)\big\rangle = \Plin(k) \biggl\{
    \sigma_{L0}(\los) \Bigl[
        \frac{16}{99}f^3 - \frac{16}{693}f^3 \Dlin \Bigl] \\
    + \sigma_{L2}(\los) \Bigl[
        \Bigl(\frac{592}{1617}f^2 + \frac{32}{99}f^3\Bigr)
        - \Bigl(\frac{16}{231}f^2 + \frac{32}{693}f^3\Bigr) \Dlin \Bigl] \biggr\}.
\end{multline}
Therefore, interestingly, at leading order the long modes give rise to a
local 6th order power spectrum multipole, even though the linear RSD itself
does not.

One subtlety arises when estimating galaxy clustering from data, for which one
needs to obtain the reference mean galaxy density from the survey itself.
In \eqref{P_mu} and \eqref{P_l}, the normalization factor $V_2=\int_\vx
\nbar^2(\vx) w^2(\vx)$ carries this reference where $\nbar$ is measured in
redshift space from the observed galaxies.
In the simplest case we assume that the selection function is overall rescaled
by the long-wavelength redshift-space galaxy overdensity
$\nbar\to\nbar(1+\Delta_\gal)$, while the change in the galaxy weights $w$ is
negligible\footnote{This is the case for our numerical test in Sec.~\ref{sec:mock},
whereas more generally $w$ also changes with $\Delta_\gal$, e.g.\ $\nbar w \to
\nbar w (1+w\Delta_\gal)$ for FKP weighting leading to relations more complicated
than \eqref{V2hat}.}.
Compared to the true $V_2$ in \eqref{V2}, the estimated normalization factor
\begin{equation}
    \label{V2hat}
    \hat V_2 = \int_\vx \WS^2(\vx) \bigl(1+\Delta_\gal(\vx)\bigr)^2
    \simeq V_2 \Bigl[1 + \bigl(2b_1+\frac23f\bigr)\Delta_0 + \frac43f\Delta_2\Bigr].
\end{equation}
We have used the fact that the above integral is dominated on large scale due
to the window function, so $\Delta_\gal\ll1$ and linear RSD applies.
Thus apart from the physical responses due to growth and dilation effects in
\eqref{Rmu} and \eqref{Rl}, the total response functions receive suppression as
a result of normalization in the estimators
\begin{align}
    \label{Rmu_sup}
    R_L(k,\mu) &= G_L + D_L \D
    - \bigl(2b_1+\frac23f\bigr)\delta_{L0} - \frac43f\delta_{L2}, \\
    \label{Rl_sup}
    R_{lL}(k) &= G_{lL} + D_{lL} \D
    - \bigl(2b_1+\frac23f\bigr)\delta_{L0} - \frac43f\delta_{L2}.
\end{align}
Together with the growth and dilation coefficients in \eqref{GD_mu} \& \eqref{GD_mu2} and
Tabs.~\ref{tab:G_l} \& \ref{tab:D_l}, the above equations are our main results
on the redshift-space galaxy response functions.

\begin{figure}[tbp]
    \centering
    \includegraphics[width=0.49\textwidth]{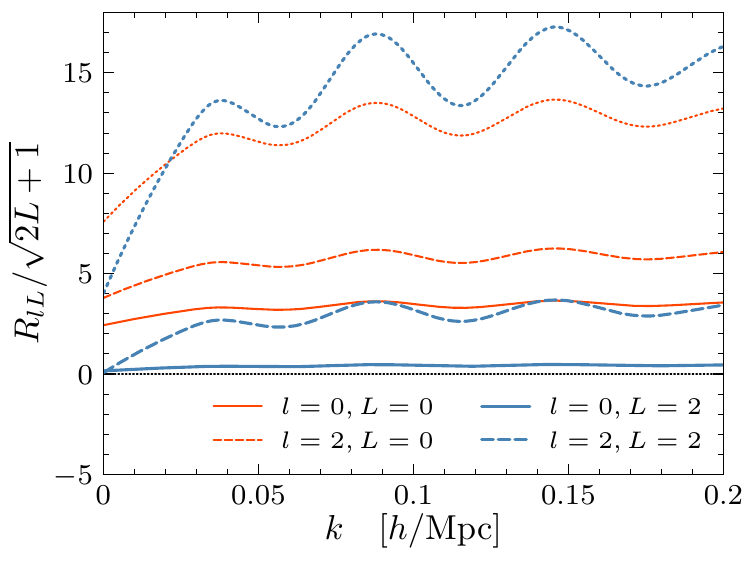}
    \includegraphics[width=0.49\textwidth]{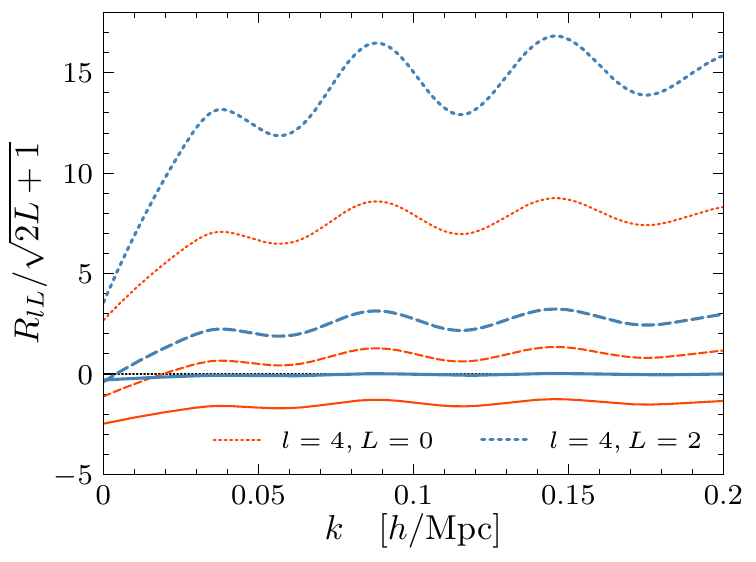}
    \caption{Redshift-space galaxy power spectrum response functions from perturbation theory.
    The left panel shows the responses only from the physical growth and
    dilation effects, while the right panel also includes the suppression due
    to normalization in the power spectrum estimator.
    The tidal responses are normalized by $\sqrt 5$ so that the modulating
    modes $\sqrt L\Delta_L$ have the same variance for an isotropic window.
    We use the following
    values $f=0.757$, $b_1=2.2$, $b_2=b_{s^2}=0$, and a dilation template
    measured from QPM mock catalogs as detailed in Sec.~\ref{sec:mock}.}
    \label{fig:Rl_ana}
\end{figure}

In Fig.~\ref{fig:Rl_ana} we show the multipole representation of the response
functions, with and without the suppression factors.
The suppression changes $R_{00}$ from $\sim3$ to $\sim-2$ and reverses its
sign, meaning for the right combination of $b_1$ and $f$ the effect of $R_{00}$
can be minimized.
Note that we have divided the tidal responses by $\sqrt5$ for comparison with
the density responses, given that for an isotropic window the rms mean tide is
$1/\sqrt5$ times the rms mean overdensity as shown in \eqref{long_cov_iso}.

\subsection{Redshift-space dilation}
\label{sub:dilation}

We can rewrite the linear perturbations in the redshift-space galaxy power
spectrum $P(k,\mu)$ due to $\Delta_0$ and $\Delta_2$ as
\begin{align}
    \Pdimless^W(k,\mu) &\simeq \Pdimless(k,\mu)
        \bigl( 1 + G_0\Delta_0 + G_2\Delta_2 \bigr)
        + \frac{\partial \Pdimless(k,\mu)}{\partial\ln k}
            \bigl( D_0\Delta_0 + D_2\Delta_2 \bigr) \nonumber\\
    &\simeq \Pdimless\bigl(k / \alpha(\mu), \mu\bigr)
        + \Pdimless(k,\mu) \bigl( G_0\Delta_0 + G_2\Delta_2 \bigr),
\end{align}
where
\begin{equation}
    \alpha(\mu)\simeq1 - D_0(\mu)\Delta_0 - D_2(\mu)\Delta_2
    \label{dilation}
\end{equation}
is determined by the dilation terms.
Obviously this shifts any feature in the power spectrum from its global
average $\langle\alpha(\mu)\rangle=1$, which is exactly the reason why the
dilation effect is named so.
In the above model we have ignored the impact on $\alpha(\mu)$ by the AP
effect, which is caused by the distortion due to the assumption of a wrong
fiducial cosmology.
It is well known that $\Delta_0$ induces isotropic dilation of
$(1+\Delta_0)^\frac13$ in real space, which follows immediately from mass
conservation.
On top of this, we expect the long-wavelength tidal perturbation and
redshift-space distortion would generate anisotropic dilation, and enhance the
isotropic one as well.

Taking $\mu=0,1$, we obtain the shift in the transverse and radial directions
\begin{align}
    \delta\alpha_\perp &\simeq \frac13 (\Delta_0 - \Delta_2), \nonumber\\
    \delta\alpha_\parallel &\simeq \frac13 (1+f) (\Delta_0 + 2\Delta_2).
    \label{alpha_ani}
\end{align}
Therefore the dilation of the volume-averaged distance $\alpha =
\alpha_\parallel^\frac13 \alpha_\perp^\frac23$ is
\begin{equation}
    \delta\alpha
    \simeq \frac13\delta\alpha_\parallel + \frac23\delta\alpha_\perp
    = \frac13\Delta_0 + \frac{f}9 (\Delta_0 + 2\Delta_2),
    \label{alpha_iso}
\end{equation}
which one can also derive by averaging $\alpha(\mu)$ over $\mu$ (taking the
monopole).
If one takes $f\to0$ and $\Delta_2\to0$, \eqref{alpha_ani} and
\eqref{alpha_iso} fall back to the well known case of real-space isotropic
dilation.
The anisotropic component is conventionally described by $\epsilon \equiv
(\alpha_\parallel/\alpha_\perp)^\frac13 - 1$, with
\begin{equation}
    \delta\epsilon
    \simeq \frac13\bigl(\delta\alpha_\parallel-\delta\alpha_\perp\bigr)
    = \frac13\Delta_2 + \frac{f}9 (\Delta_0+2\Delta_2),
    \label{epsilon}
\end{equation}
which is really half the quadrupole moment of $\alpha(\mu)$.
As expected it can arise from either $\Delta_2$ or $f$.
$\alpha$ and $\epsilon$ constitute an alternative parametrization of
$\alpha_\perp$ and $\alpha_\parallel$ up to quadrupole order.

Importantly, the anisotropic redshift-space dilation
 breaks the assumption of the AP
test that any feature be
statistically isotropic provided the fiducial cosmology is the true one.
Additionally, the redshift-space dilation also shifts the BAO peak
anisotropically.
Unlike the shift of the peak due to BAO-scale nonlinear structure formation,
this effect is intrinsic to a survey itself and cannot be removed by
reconstruction methods without knowing the amplitudes of the long modes beyond
scales probed by the survey.
We will quantify the impact of super-survey modes on the cosmological
constraints from both BAO fitting and full power spectrum shape analysis in
Sec.~\ref{sub:impact}.

\section{Responses in galaxy mock catalogs}
\label{sec:mock}

Sec.~\ref{sub:mock_resp} describes the galaxy mocks and the estimators to
measure response functions from those mocks.
In Sec.~\ref{sub:mock_results} we compare these numerically calibrated responses
to the analytic one from Sec.~\ref{sec:ana}.

\subsection{QPM mock responses}
\label{sub:mock_resp}

The quick particle mesh (QPM) \cite{White14b} method is a fast algorithm to
generate galaxy mocks that match abundance and clustering properties.
This method selects a subset of PM simulation particles and elevates them as mock
dark matter halos based on their local density smoothed on some nonlinear
scale.
To measure the responses, we exploit $N_\mock=991$ QPM mocks of size
$2560\Mpch$ originally produced for the BOSS DR12 analysis \cite{AlamEtAl17}.
We use the mock fiducial values $f=0.757$, $b_1=2.2$, and take $b_2=b_{s^2}=0$
given the QPM algorithm only traces the density.

For each QPM mock we use the galaxy catalog from the full periodic box without
the survey mask, and instead sample the short modes by picking galaxies from
$N_\sub=64$ sub-volumes, each having some nonzero long modes.
For robustness we adopt two types of sub-volumes of different geometries and
volumes: tiling sub-boxes of size $640\Mpch$, and sub-spheres of radius
$320\sqrt3\Mpch$ centered on the sub-box vertices.
In the following analysis, we always use a $512^3$ grid to perform the Fast
Fourier Transformation (FFT), and the cubic order scheme to paint galaxies onto
the grid \cite{SefusattiEtAl16}.

To measure the amplitudes of the long modes, we can simply follow \eqref{long}
and multiply the galaxy field in Fourier space by the sub-volume window and
Legendre polynomial, before tranforming backward to read off $b_1\Delta_L$'s at
their corresponding vertex positions in configuration space.
By further dividing by $b_1$, we can remove the large scale galaxy biasing to
obtain the long modes in matter density.
%
%
Recall that $\Delta_0$ and $\Delta_2$ are defined in real space, and the latter
depends on the LOS direction $\los$.
With so many sub-volumes we can estimate the variance of the long modes
$\sigma_{LL}$ accurately.

Next let's estimate the short-mode power spectrum multipoles $\hat P_l^W$ in
redshift space.
To improve statistics, for each sub-box we apply RSD by adding peculiar
velocities in 6 LOS's -- 3 principle axes and 3 face diagonals, allowing some
redundancy given there are only 5 independent components of $\Delta_2$.
Then we paint the masked full volume onto the FFT grid, before subtracting off
the mean using random catalogs (50 times as abundant) inside the sub-volumes
and padding with zeros outside.
We take advantage of the interlacing technique, which together with the cubic
order assignment scheme greatly suppresses aliasing \cite{SefusattiEtAl16}.
We then deconvolve the latter \cite{Jing05}, subtract the constant shot noise
from the squared field, and finally bin the power spectrum multipoles in
spherical shells of width $0.01\hMpc$.

Now we can compute the redshift space response functions by correlating
$\Delta_L$ and $\hat P_l^W$ pairs, with a total of $N_\tot=6N_\mock
N_\sub=380544$ realizations,
\begin{equation}
    \hat R_{lL}(k) = \frac{\sum_i^{N_\tot} \hat P_{li}^W\!(k)\, \Delta_{Li}}
                        {N_\tot\, P_l^W(k)\, \sigma_{LL}}
    = \frac{N_\tot \sum_i^{N_\tot} \hat P_{li}^W\!(k)\, \Delta_{Li}}
                        {\sum_i^{N_\tot} \hat P_{li}^W(k) \sum_j^{N_\tot} \Delta_{Lj}^2}
\end{equation}
The error is obtained by bootstrapping within all mocks.
There are 2 options for the reference number density when measuring $\hat P_l^W$:
the average of all mocks or the sub-volume mean.  Using the former we will get
the response entirely from the physical growth and dilation effect \eqref{Rl},
while taking the latter we also include the suppression due to local
normalization \eqref{Rl_sup}
To relate to the previous works, the former case is named ``global'' whereas
the latter is referred to as ``local'' \cite{LiHuEtAl14}.

\subsection{Comparison to perturbation theory}
\label{sub:mock_results}

\begin{figure}[tbp]
    \centering
    \includegraphics[width=\textwidth]{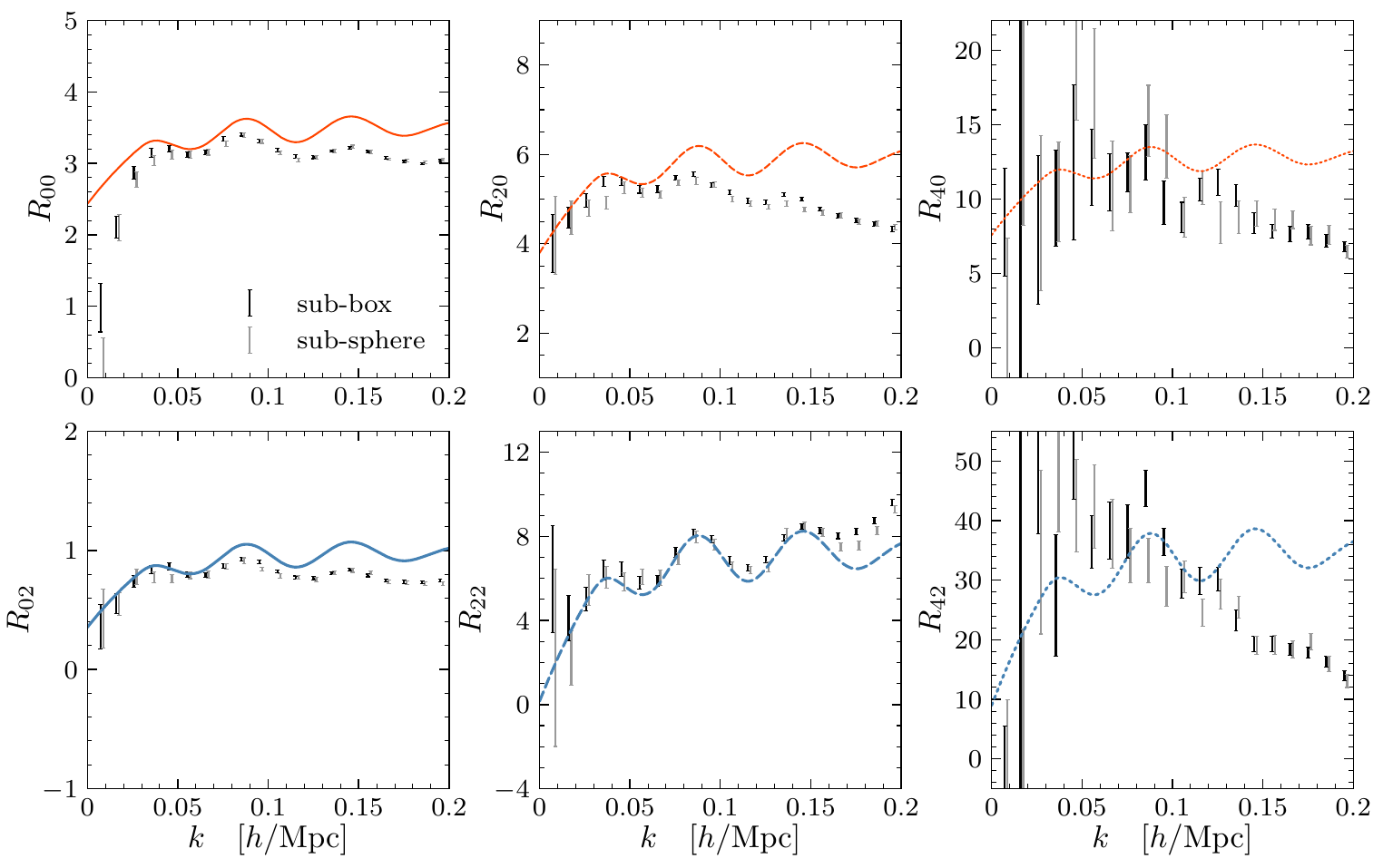}
    \caption{Redshift-space response functions from galaxy mocks due to
    physical growth and dilation effects.  Black and grey points with error
    bars are the mock measurements from sub-boxes and sub-spheres
    respectively.  Lines shows the perturbation theory results.}
    \label{fig:Rl_glb}
\end{figure}

\begin{figure}[tbp]
    \centering
    \includegraphics[width=\textwidth]{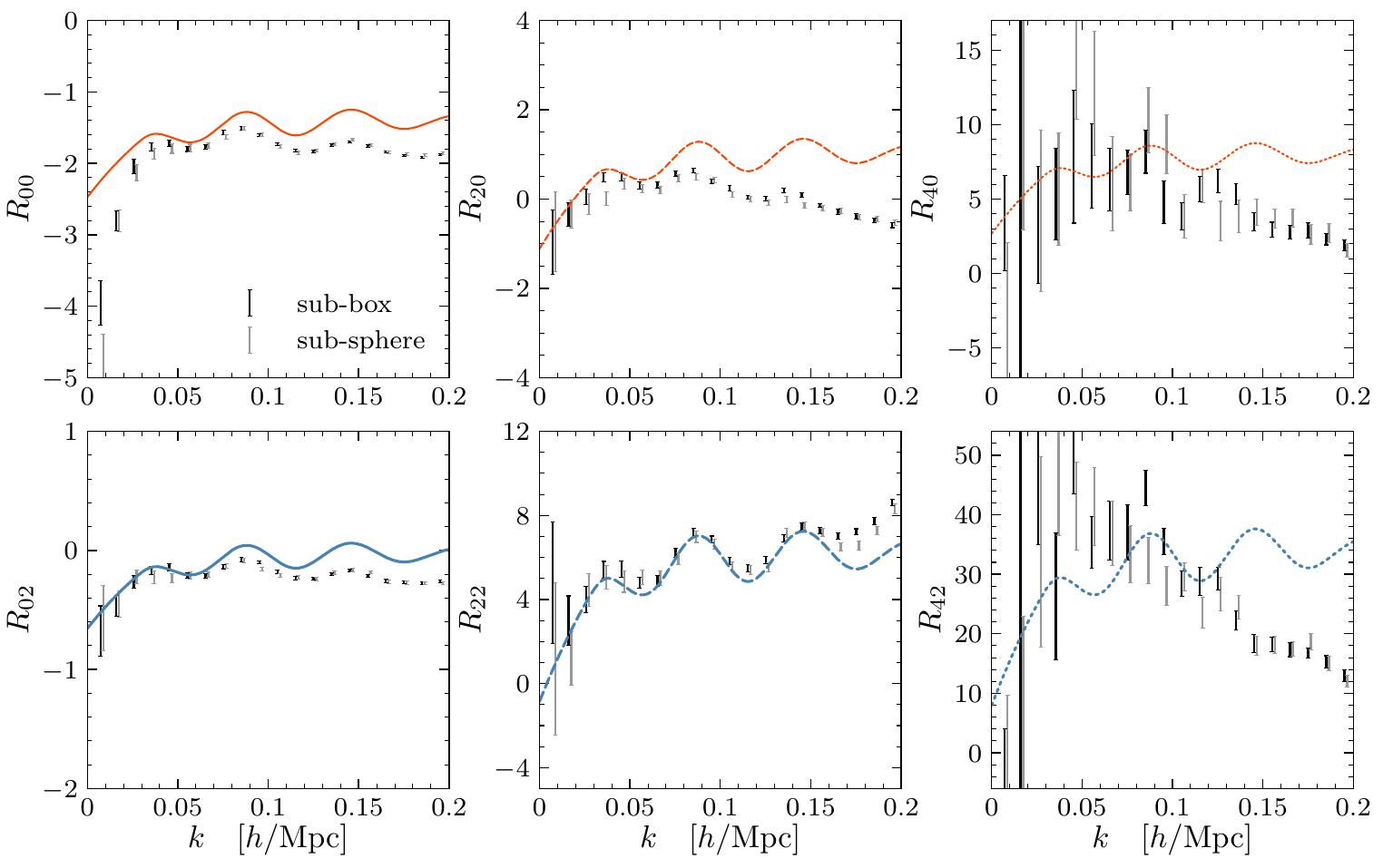}
    \caption{Same as Fig.~\ref{fig:Rl_glb}, except that the responses also
    include suppression resulting from normalization of the estimator.  The
    suppression is about $-5$ for density responses and $-1$ for tidal
    responses.}
    \label{fig:Rl_loc}
\end{figure}

Before comparing the analytic predictions in Sec.~\ref{sec:ana} to the mock
response functions, one still needs a dilation template $\d\ln\Pdimless/\d\ln
k$.
We measure this by taking the derivative of the cubic-splined average power
spectrum of all the real-space full boxes.

Fig.~\ref{fig:Rl_glb} shows the comparison of mock and analytic redshift-space
response functions due to growth and dilation effects, without suppression from
the normalization factor of the estimator.
The sub-box and sub-sphere results agree with each other throughout the range
of $k$, and they are both consistent with the perturbative responses on
large scales up to $0.06\hMpc$, beyond which the agreement worsens due to
nonlinearity.
Here the only exception is that the two mock $R_{00}$'s have some sharp
downturn at $k\lesssim0.03\hMpc$.
Given that this scale-dependence is too sharp, and it only occurs in the
monopole response to mean density, this feature does not seem to be physical.
And indeed the QPM algorithm selects particles to stand in for halos based on
the smoothed density, therefore the smoothing scale results in some
mode-coupling leading to a large scale scale-dependent bias, which could
possibly give rise to such an artifact.

In Fig.~\ref{fig:Rl_loc} we show the same comparison, but in addition to the
physical responses we also include the suppression resulting from the
normalization factor in the power spectrum estimator as in \eqref{Rl_sup}.
The agreement between the analytic and numerical results is as good as in
Fig.~\ref{fig:Rl_glb}.
Including the suppression from the estimator normalization leads to an offset by
 about $-5$ for the density responses and
approximately $-1$ for the tidal responses.
As a result, the density response $R_{00}$ of the power spectrum monopole has
its sign reversed, so that its effect could be minimized for the right value
of $b_1$ and $f$.

\section{Super-sample effect}
\label{sec:ss}

In Sec.~\ref{sub:ssc} we briefly review  the power spectrum
covariance, especially the super-sample covariance due to the presence of long
modes, before comparing the SSC model to the one measured from mocks in
Sec.~\ref{sub:cov}.
Finally in Sec.~\ref{sub:impact} we assess the impact of this extra covariance
on the BAO, AP, and RSD measurements.

\subsection{Redshift space super-sample covariance}
\label{sub:ssc}

Now let's consider a galaxy survey with size and geometry described by its
window function $W$.
The statistical precision of power spectrum measurement is described by its
covariance.
The long modes beyond the size of the survey correlate the observed power
spectrum at all wavelengths, giving rise to an additional error known as the
super-sample covariance.
The covariance of the galaxy power spectrum is defined by
\begin{equation}
    C(\vk_1,\vk_2) \equiv \bigl\langle \bigl(\hat P^W(\vk_1)-P^W(\vk_1)\bigr)
    \bigl(\hat P^W(\vk_2)-P^W(\vk_2)\bigr) \bigr\rangle,
\end{equation}
which can be further decomposed into three pieces based on their statistical and
physical origin \cite{TakadaHu13}
\begin{equation}
    \label{C}
    C(\vk_1,\vk_2) = C^\Gauss(\vk_1,\vk_2) + C^{T0}(\vk_1,\vk_2) + C^\SS(\vk_1,\vk_2).
\end{equation}
The first term is the only nonvanishing piece for a Gaussian random field, and
hence its superscript.
Therefore we refer the other two terms jointly as the non-Gaussian covariance,
in which the second piece arises from direct mode coupling between scales
$\vk_1$ and $\vk_2$, and the last piece or the super-sample covariance
originates from the indirect mode coupling through the coherent modulation by
the long modes.

In the small-scale limit ($k_1, k_2\gg V^{-\frac13}$) the covariance
expressions simplify.
The Gaussian covariance becomes diagonal
\begin{equation}
    C^\Gauss(\vk_1,\vk_2) \simeq \frac{2(2\pi)^3}{V_\eff} \deltaD(\vk_1 - \vk_2)
    \Bigl( P(\vk_1) + \frac1{\nbar} \Bigr)^2,
\end{equation}
which includes the contribution of a constant shot noise.
We have defined an effective volume as
\begin{equation}
    V_\eff = \frac{V_2^2}{V_4},
\end{equation}
that differs from that in Ref.~\cite{Tegmark97gal} in that the former only
manifestly includes the volume effect and is not explicitly sensitive to the
level of shot noise.
And following the notation of Ref.~\cite{TakadaHu13} the second term $C^{T0}$,
named the trispectrum piece, is given by
\begin{equation}
    \label{C_T0}
    C^{T0}(\vk_1,\vk_2) \simeq \frac1{V_\eff} T(\vk_1,-\vk_1,\vk_2,-\vk_2),
\end{equation}
where $T$ is the redshift-space galaxy trispectrum or the connected 4-point
function
\begin{equation}
    \bigl\langle\delta(\vk_1)\delta(\vk_2)\delta(\vk_3)\delta(\vk_4)\bigr\rangle_\mathrm{c}
    = (2\pi)^3 \deltaD\bigl({\textstyle\sum_{i=1}^4}\vk_i\bigr)\,
    T(\vk_1,\vk_2,\vk_3,\vk_4).
\end{equation}
Lastly, the super-sample covariance reads intuitively
\begin{equation}
    C^\SS(\vk_1,\vk_2) = \sum_{L_1L_2} \sigma_{L_1L_2}
    \frac{\d P^W(\vk_1)}{\d\Delta_{L_1}}\frac{\d P^W(\vk_2)}{\d\Delta_{L_2}}.
\end{equation}

In practice one bins the power spectrum in shells of finite width, so that its
covariance function reduces to a matrix.
Specifically we want to evaluate the Gaussian covariance matrix, which we can
subtract from the full covariance matrix to obtain the non-Gaussian piece.
Taking the multipole moments in the spherical $\vk$-shells labeled by $k_i,k_j$
\cite{GriebEtAl16}, we get\footnote{Our equations differ from those in
Ref.~\cite{GriebEtAl16} by a factor of $(2l_5+1)$.}
\begin{multline}
    \label{Cll_G}
    C^\Gauss_{l_1l_2}(k_i, k_j) \equiv (2l_1+1)(2l_2+1)
    \int_{\substack{\vspace{1ex}\\ \vk_1\in V_{k_i} \\ \vk_2\in V_{k_j}}}
    C^\Gauss(\vk_1, \vk_2) \L_{l_1}(\mu_1)\L_{l_2}(\mu_2) \\
    = \frac{2(2l_1+1)(2l_2+1)\delta_{ij}}{N_{k_i}}
    \sum_{l_3l_4l_5} (2l_5+1)
    \begin{pmatrix} l_1 & l_2 & l_5 \\ 0 & 0 & 0 \end{pmatrix}^2
    \begin{pmatrix} l_3 & l_4 & l_5 \\ 0 & 0 & 0 \end{pmatrix}^2
    \int_{\vk\in V_{k_i}} P_{l_3}(k)P_{l_4}(k)
\end{multline}
where the big parentheses are the Wigner 3-j symbols and $N_{k_i} \equiv V_\eff
V_{k_i}/(2\pi)^3$ is the number of modes in the $i$th shell.
The triple sum over $l_3,l_4,l_5$ runs from 0 to infinity for each $l$, and the
summands are nonzero only when the following selection rules are satisfied:
$\max(|l_1-l_2|,|l_3-l_4|)\leq l_5\leq\min(l_1+l_2,l_3+l_4)$, and $l_5$ should
have the same parity as $l_1+l_2$ and $l_3+l_4$ (i.e., even).
Note that the shot noise term only shows up with the power spectrum monopoles,
and is left implicit here.
Similarly the trispectrum piece becomes
\begin{equation}
    \label{Cll_T0}
    C_{l_1l_2}^{T0}(k_i, k_j) = \frac1{V_\eff} T_{l_1l_2}(k_i,k_j),
\end{equation}
where
\begin{equation}
    T_{l_1l_2}(k_i,k_j) \equiv (2l_1+1)(2l_2+1)
    \int_{\substack{\vspace{1ex}\\ \vk_1\in V_{k_i} \\ \vk_2\in V_{k_j}}}
    T(\vk_1,-\vk_1,\vk_2,-\vk_2) \L_{l_1}(\mu_1)\L_{l_2}(\mu_2).
\end{equation}
The super-sample covariance is simply
\begin{equation}
    \label{Cll_SS}
    C_{l_1l_2}^\SS(k_i, k_j) =
    P_{l_1}(k_i) P_{l_2}(k_j) \sum_{L_1L_2} \sigma_{L_1L_2}
    R_{l_1L_1}(k_i) R_{l_2L_2}(k_j).
\end{equation}

Eqs.~\eqref{Cll_G}, \eqref{Cll_T0}, and \eqref{Cll_SS} show clearly how
different covariance pieces scale with survey volume and binning.
Both $C^\Gauss$ and $C^{T0}$ are inversely proportional to $V_\eff$ with the
former depending in addition on binning, whereas $C^\SS$ scales as the
covariance of the long modes $\sigma_{L_1L_2}$.
From \eqref{long_cov}, we can see that $\sigma_{L_1L_2}$ roughly scales as
inverse volume when the window size is near the peak of the power spectrum, in
which case all three covariance pieces would have the same volume scaling.
Increasing the volume beyond that makes $C^\SS$ decrease more rapidly than the
other two terms.
However, this difference in volume scaling is very small, e.g.,
$V\sigma_{00}(V)$ only decreases by a factor of 2 when a cubic volume $V$
increases from $0.1\,\Gpc^3/h^3$ to $10\,\Gpc^3/h^3$ by 2 orders of magnitude.

In the above equations we have omitted the shot noise terms
\cite{MeiksinWhite99} other than the one included in the Gaussian piece.
For sufficiently wide bins that have more modes than the number of galaxies $N_\gal$,
those off-diagonal shot noise terms start to become important, with the leading order
being $\order(B/N_\gal)$, where $B$ refers to the bispectrum terms in Eq.8 of
Ref.~\cite{MeiksinWhite99}.
Higher order terms including $\order(P/\nbar N_\gal)$ and
$\order(1/\nbar^2 N_\gal)$ kick in as well when $\nbar P\lesssim1$.

\subsection{Mock covariance}
\label{sub:cov}

Now let's examine if the redshift galaxy response functions give rise to
the super-sample covariance expected from \eqref{Cll_SS}, using again the QPM
mocks described in Sec.~\ref{sub:mock_resp}.
If we measure the covariance matrix from the $N_\mock$ periodic boxes, the
result would only contain the Gaussian and trispectrum contribution since all
modes beyond the box size vanish.
For each of the 6 LOS, we estimate the covariance matrix among all full boxes
\begin{equation}
    \label{C_los_est}
    \hat C_{l_1l_2}(k_i,k_j;\los) =
    \frac1{N_\mock-1}\sum_{b=1}^{N_\mock}
    \bigl(\hat P_{l_1b}(k_i)-P_{l_1}(k_i)\bigr)
    \bigl(\hat P_{l_2b}(k_j)-P_{l_2}(k_j)\bigr).
\end{equation}
before averaging over all 6 LOS's for a combined estimation
\begin{equation}
    \label{C_est}
    \hat C_{l_1l_2}(k_i,k_j) = \frac16 \sum_\los \hat C_{l_1l_2}(k_i,k_j;\los).
\end{equation}

Given the Gaussian covariance is completely determined by the power spectrum,
we can evaluate it using the full box power spectrum multipoles as specified in
\eqref{Cll_G}.
Because $C^\Gauss$ is sensitive to the choice of binning, for a clean
comparison we would like to subtract it off from \eqref{C_est} to obtain the
non-Gaussian covariance $\hat C^\nonGauss$ that only depends on survey volume
and geometry.
In the case of periodic boxes, $\hat C^\nonGauss$ includes the trispectrum
piece $C^{T0}_{l_1l_2}$ and off-diagonal shot noise terms.
So it serves as a baseline for comparison with the full covariance matrix
including SSC, the latter to be measured from the sub-volumes.
Because of the difference in volume between the periodic box and sub-volumes,
we need to rescale the periodic covariance by a volume factor of $V/V_\sub$
before comparing the two.

The sub-volumes have nonzero long modes above their size, so their covariance
contains the SSC piece in $C^\nonGauss$.
To measure it, we can use a estimator similar to the one in \eqref{C_los_est}
and \eqref{C_est}.
Because sub-volumes from different mocks are independent, for each LOS and each
sub-volume location labeled with $s$, the covariance estimator is
\begin{equation}
    \label{Csub_los_est}
    \hat C_{l_1l_2}(k_i,k_j;\los,s) =
    \frac1{N_\mock-1}\sum_{b=1}^{N_\mock}
    \bigl(\hat P^W_{l_1b}(k_i)-P^W_{l_1}(k_i)\bigr)
    \bigl(\hat P^W_{l_2b}(k_j)-P^W_{l_2}(k_j)\bigr).
\end{equation}
And then we further average over the LOS's and sub-volume locations
\begin{equation}
    \label{Csub_est}
    \hat C_{l_1l_2}(k_i,k_j)
    = \frac1{6N_\sub} \sum_{\los,s} \hat C_{l_1l_2}(k_i,k_j;\los,s).
\end{equation}
We estimate the error on \eqref{C_est} and \eqref{Csub_est} by bootstrapping
over all $N_\mock$ boxes.
The masked covariance matrix is affected by the window function and leaks
off-diagonally, which can be suppressed in the small-scale wide-bin limit.
For simplicity we subtract the volume-scaled diagonal Gaussian covariance
\eqref{Cll_G} assuming no window effect.
Because of this we expect to underestimate the sub-volume diagonal
$C^\nonGauss$ and slightly overestimate its immediate off-diagonal elements.

To verify our SSC model in \eqref{Cll_SS}, we can add it to the volume-scaled
periodic covariance to see if this reproduces the sub-volume covariance.
To evaluate it we use the mean power spectrum multipoles of the periodic boxes,
the mock response functions measured in Sec.~\ref{sub:mock_resp}, and the
covariance of the long modes $\sigma_{L_1L_2}$ estimated by averaging
$\Delta_{L_1}\Delta_{L_2}$ over all sub-volumes and all LOS's.

\begin{figure}[tbp]
    \centering
    \includegraphics[width=\textwidth]{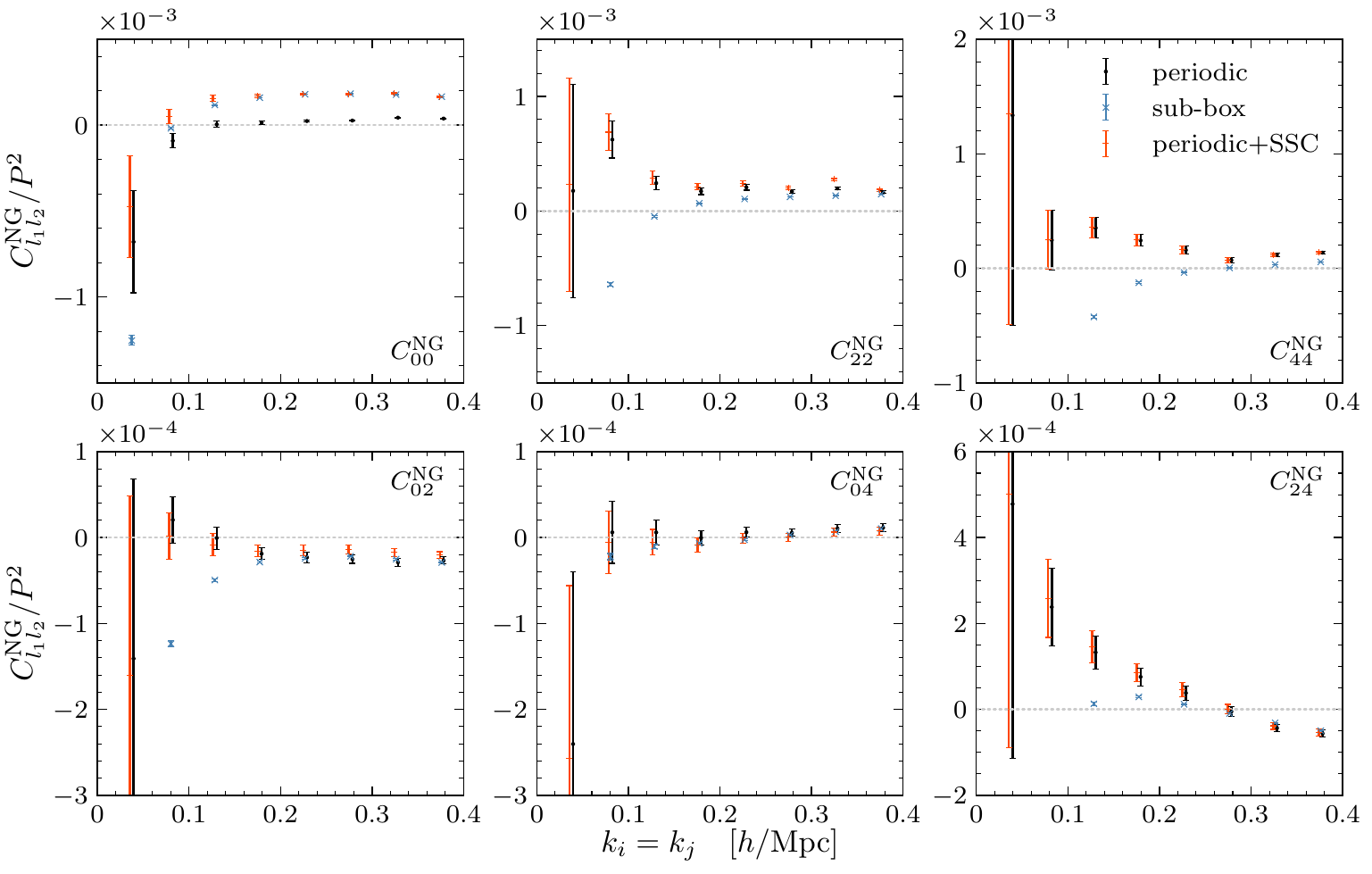}
    \caption{Redshift-space non-Gaussian covariance 
    $C^\mathrm{NG}_{l_1l_2}=\mathrm{cov}(P_{l_1}(k_i),P_{l_2}(k_j))$ 
    of galaxy power spectrum
    multipoles, for diagonal elements $k_i=k_j$.
    The covariance is divided by the squared real-space power
    spectrum.
    }
    \label{fig:Cl_Kdiag}
\end{figure}

Fig.~\ref{fig:Cl_Kdiag} shows the non-Gaussian diagonal elements of the auto-
and cross-covariance matrices of power spectrum multipoles ($l=0,2,4$) measured
from the QPM mocks.
We have normalized $C^\nonGauss$ by the real-space power spectrum
$P(k_i)P(k_j)$ for a dimensionless quantity.
As expected we notice that the sub-box $C^\nonGauss$ turns negative on large
scale due to an over-subtraction of the Gaussian contribution, and more so for
larger $l$'s.
We have also chosen a wide bin width of $\Delta k=0.05\hMpc$ to suppress
the window effect.
Clearly the biggest contribution of SSC is to the monopole auto-covariance
$C^\nonGauss_{00}$, in which it dominates over the rest by a factor of
$5\sim10$, where there's enough statistical precision ($k\gtrsim0.15\hMpc$).
We find that \eqref{Cll_SS} models very well the SSC as the difference between periodic
and sub-volume covariance.
However, SSC does not add significantly to the other $C^\nonGauss_{l_1l_2}$'s.
This is because not only the SSC becomes smaller for $l>0$, but also the other
non-Gaussian terms (trispectrum and shot noise contribution) become much
larger.
As for the latter cause, the trispectrum term should dominate over the shot
noise on large scales where $\nbar P\gg1$.
Therefore the fact that $C^\nonGauss_{22}$ and $C^\nonGauss_{44}$ are much
bigger than the periodic $C^\nonGauss_{00}$ at $k\approx0.1\hMpc$ is hinting at
some strong angular dependence in the redshift-space galaxy trispectrum
\eqref{C_T0}.

\begin{figure}[tbp]
    \centering
    \includegraphics[width=\textwidth]{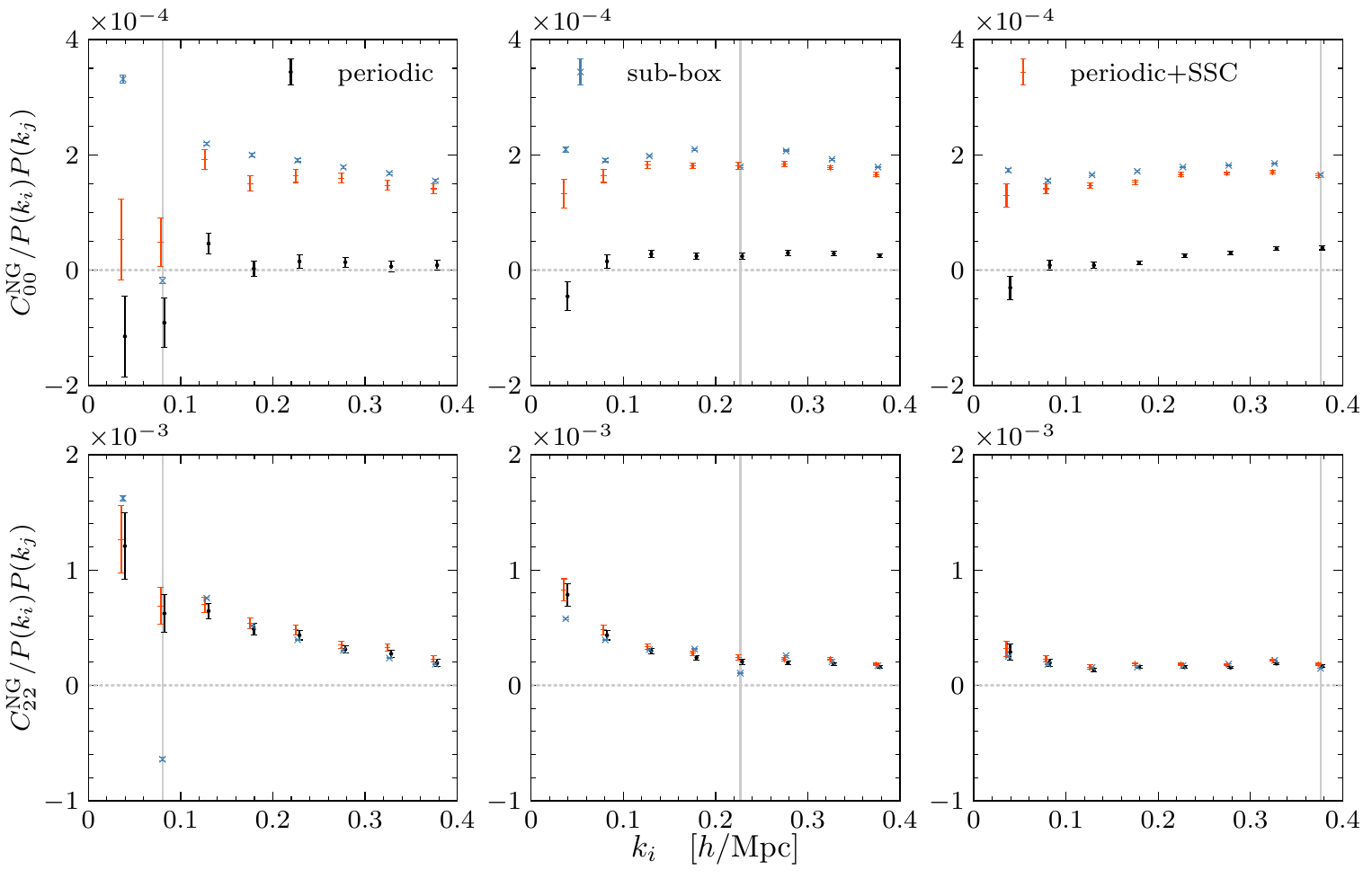}
    \caption{Off-diagonal redshift-space non-Gaussian covariance
      $C^\mathrm{NG}_{ll}=\mathrm{cov}(P_{l}(k_i),P_{l}(k_j))$ 
    of the galaxy power spectrum
    monopole and quadrupole, evaluated as a function of $k_i$ for different 
    fixed $k_j$'s marked by vertical lines.
    The covariance is divided by the squared real-space power
    spectrum.
    }
    \label{fig:Cl_Koffd}
\end{figure}

In Fig.~\ref{fig:Cl_Koffd} we show the off-diagonal elements of the
auto-covariance matrices of power spectrum multipoles ($l=0,2$) at 3 fixed
$k_j$'s, binned and normalized in the same way as in Fig.~\ref{fig:Cl_Kdiag}.
Again because of the leakage due to the window function, the sub-box covariance
is underestimated at $k_i=k_j$ and overestimated in the neighboring bins,
especially on the large scale.
Similar to Fig.~\ref{fig:Cl_Kdiag}, SSC dominates $C^\nonGauss_{00}$ by a
factor of $5\sim10$ in its off-diagonal elements, but is negligible in all
other $C^\nonGauss_{l_1l_2}$'s, for which only $C^\nonGauss_{22}$ is shown here
given their qualitative similarity.
The three cases of $C^\nonGauss_{22}$ overlap well with each other, and are
about one order of magnitude larger than the periodic $C^\nonGauss_{00}$.
This again implies some interesting angular dependence of the trispectrum term
\eqref{C_T0}, which is worth investigating further but is beyond the scope of this
paper.
Figs.~\ref{fig:Cl_Kdiag} and \ref{fig:Cl_Koffd} present the results for the
sub-boxes, and we get qualitatively the same results with sub-spheres.

\subsection{Impact on BAO, AP, and RSD constraints}
\label{sub:impact}

\begin{figure}[tbp]
    \centering
    \includegraphics[width=\textwidth]{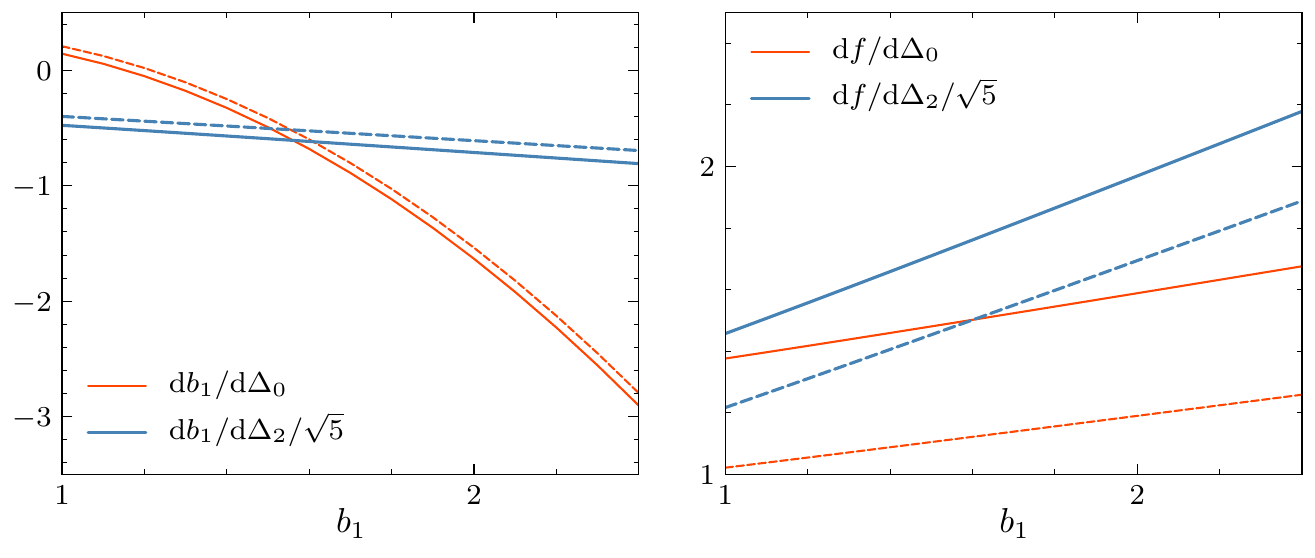}
    \caption{Response of error in $b_1$ and $f$ induced by the long modes.
    We have assumed $f=0.9$(solid), $0.7$(dashed), $b_2=0$, $\d\ln\Pdimless /
    \d\ln k\approx1.5$, $b_{s^2}=-4(b_1-1)/7$.  The tidal responses are
    normalized by $\sqrt 5$ so that the modulating modes $\sqrt L\Delta_L$ have the same variance for an
    isotropic window.}
    \label{fig:bf_resp}
\end{figure}

Because the super-survey modes generate an excess galaxy power spectrum
covariance, constraints on cosmological parameters will be degraded due to
their stochasticity.
In this paper we are interested in the degradation of constraints from
spectroscopic galaxy surveys, namely from BAO, AP, and RSD measurements.
To evaluate this, a typical approach is the Fisher information analysis (see e.g.\
\cite{LiHuEtAl14sss}).
Here we adopt a much simpler and yet equivalently effective approach in three
separate steps.
We will first demonstrate how to translate the amplitudes of $\Delta_L$ to
systematic errors on various parameters such as the BAO scale $\alpha$ and growth rate $f$, 
then show how
to assess the size of the effect given the window function using the BOSS survey as an
example, and finally compare that to the BOSS DR12 constraint.

The presence of long modes gives rise to systematic errors in BAO and AP
straightforwardly according to \eqref{alpha_ani}, \eqref{alpha_iso}, and
\eqref{epsilon}.
The errors on those geometry parameters will further contaminate the
conventionally quoted (e.g.\ \cite{AlamEtAl17, BeutlerEtAl17}) physical
quantities, i.e.\ the angular diameter distance $\Dist{A}(z)$, the Hubble
expansion rate $H(z)$, the spherically-averaged distance $\Dist{V}(z)$, and the
AP parameter $\FAP(z)$.
The anisotropic BAO measurement constrains $\Dist{A}(z)$ and $H(z)$, whose
systematic errors translate as
\begin{equation}
    \label{BAO_ani}
    \frac{\delta\Dist{A}}{\Dist{A}} = \delta\alpha_\perp,
    \qquad
    \frac{\delta H}H = - \delta\alpha_\parallel,
\end{equation}
whereas the isotropic BAO measurement constrains the volume-averaged distance
\begin{equation}
    \label{BAO_iso}
    \frac{\delta\Dist{V}}{\Dist{V}} = \delta\alpha.
\end{equation}
The AP measurement, for either the BAO scale or the full shape, assumes that
any isotropic feature remains isotropic when the fiducial cosmology is exactly
the true cosmology.
However this assumption is broken by the dilation caused by the
long modes, and this can be captured by the following systematic error in the
AP parameter
\begin{equation}
    \frac{\delta \FAP}{\FAP} = - 3 \delta\epsilon.
\end{equation}

To find out how an error on $f$ arises given $\Delta_0$ and $\Delta_2$, we can fit
the perturbed power spectrum multipoles with some $f$ and $b$ offset, which are
exactly the systematic error introduced when ignoring the super-sample effect.
For simplicity we use our analytic response functions \eqref{Rl_sup} and the
Kaiser formula \eqref{Kaiser} that are valid on large scale.
Assuming that most of the information is from $P_0$ and $P_2$ and that the
responses are approximately scale independent which is true around the relevant
scale, one can write down analytically
\begin{align}
    \delta P_0 &= \sum_L P_0 R_{0L} \Delta_L =
    \Plin \bigl( 2b_1\delta b_1 + \frac23b_1\delta f + \frac23f\delta b_1
        + \frac25f\delta f \bigr) \nonumber\\
    \delta P_2 &= \sum_L P_2 R_{2L} \Delta_L =
    \Plin \bigl( \frac43b_1\delta f + \frac43f\delta b_1 + \frac87f\delta f \bigr).
\end{align}
By inverting them we derive the linear responses of systematic errors with
respect to the long modes $\d b_1/\d\Delta_L$ and $\d f/\d\Delta_L$.
In practice the RSD measures the combination $b_1\sigma_8$ and $f\sigma_8$,
where $\sigma_8$ is the rms density fluctuation in a $8\Mpch$ spherical tophat.
For conciseness we do not include the $\sigma_8$ factors explicitly in the
error response expressions, but this will not affect the result.

Fig.~\ref{fig:bf_resp} shows these responses as functions of $b_1$ and $f$.
While we vary $b_1$ continuously and choose $f=0.7,0.9$, the other parameters
are held fixed $b_2=0$, $\d\ln\Pdimless/\d\ln k=1.5$, and $b_{s^2}=-4(b_1-1)/7$.
In the left panel, the error response of $b_1$ to $\Delta_2$ is sensitive to
neither $b_1$ nor $f$, while its error response to $\Delta_0$ is sensitive to
$b_1$ but not $f$.
The former and most of the latter are negative, and at higher bias
$b_1\gtrsim1.5$, the latter has a stronger effect.
Given that the linear bias is a nuisance parameter, we are mostly interested in
the error responses of $f$, which actually does not depend on the value of
$b_2$.
The right panel shows they are more sensitive to $f$ compared to the error
responses of $b_1$.
While the responses to both $\Delta_0$ and $\Delta_2$ are positive, the effect
of the latter is bigger.
In most cases, larger $b_1$ or $f$ leads to bigger effects.

\begin{table}[htpb]
    \centering
    \caption{Covariance of super-survey mode for BOSS DR12 NGC for 3 redshift
    bins, estimated using random catalogs with $N_\ran=10000$ points.}
    \label{tab:sigmaLL_boss}
    \begin{tabular}{c|ccc}
        \hline
        & $\sqrt{\sigma_{00}}$ & $\sqrt{\sigma_{22}}$ & $\rho_{02}$ \\
        \hline
        $0.2<z<0.5$ & 0.0031 & 0.0015 & 0.34 \\
        $0.4<z<0.6$ & 0.0029 & 0.0014 & 0.39 \\
        $0.5<z<0.75$& 0.0027 & 0.0013 & 0.40 \\
        \hline
    \end{tabular}
\end{table}

\begin{table}[htpb]
    \centering
    \caption{Standard deviations of the super-sample errors on various
    parameters, for the 3 redshift bins of BOSS DR12 NGC.
    For comparison, the BOSS DR12 constrains $\Dist{A}$, $H$, and $f\sigma_8$
    to $\sim1.5\%$, $\sim2\%$, and $\sim8\%$ precision, respectively, in all
    redshift bins.
    Therefore the super-sample errors are of marginal significance for BOSS.
    }
    \label{tab:impact}
    \begin{tabular}{c|cccccc}
        \hline
        & $\sigma_{\Dist{A}}/\Dist{A}$ & $\sigma_H/H$ & $\sigma_{\Dist{V}}/\Dist{V}$
        & $\sigma_{\FAP}/\FAP$ & $\sigma_{b_1\sigma_8}/b_1\sigma_8$ & $\sigma_{f\sigma_8}/f\sigma_8$ \\
        \hline
        $0.2<z<0.5$ & 0.10\% & 0.29\% & 0.14\% & 0.27\% & 0.4\% & 1.2\% \\
        $0.4<z<0.6$ & 0.09\% & 0.27\% & 0.13\% & 0.24\% & 0.3\% & 1.1\% \\
        $0.5<z<0.75$& 0.08\% & 0.26\% & 0.12\% & 0.23\% & 0.3\% & 1.0\% \\
        \hline
    \end{tabular}
\end{table}

Given the error responses, we only need the covariance of $\Delta_L$'s to
estimate the size of super-sample effects.
To compute this we apply the random pair summation algorithm presented in
App.~\ref{sec:long_cov} to the BOSS DR12 North Galactic Cap (NGC) region using
catalogs of $N_\ran=10000$ random points, for the same three redshift bins as
in the BOSS analysis \cite{AlamEtAl17}, and list the result in
Tab.~\ref{tab:sigmaLL_boss}.
There are two caveats regarding our estimation of $\sigma_{LL'}$: we have
assumed the plane parallel approximation where all pairs share a common LOS, and
have only accounted for the volume of NGC, ignoring the South Galactic Cap (SGC).
However, since the NGC sky coverage is limited and NGC is much larger
than SGC with about 3 times the number of galaxies, we do not expect drastic
difference from a more rigorous treatment.
The rms $\Delta_0$ is roughly equal to the isotropic value $\sqrt5\Delta_2$,
but slightly larger for lower redshifts.
The two super-survey modes are positively correlated, with a significantly
nonzero cross-correlation coefficient.
This implies that their impacts on the $f\sigma_8$ constraint likely add up, given that
both $\d f/\d\Delta_L$ responses are positive.
Similarly, such positive correlation will enhance the super-sample effect on
$H(z)$, $\Dist{V}(z)$, and $\FAP(z)$, while it partially cancels the
impact on $\Dist{A}(z)$.

Because $\Delta_L$'s are Gaussian modes and the error responses are linear, the
systematic errors on various parameters also follows a normal distribution
whose variance can be calculated straightforwardly.
For all three redshift bins, we combine the effect from the two degrees of
freedom and quote the fractional standard deviation $\sigma_X/X$ of the
systematic error for each parameter $X$ in Tab.~\ref{tab:impact}.
To evaluate these, we have taken $f=0.757$, $b_1=2.2$ in the responses, and
ignored their relatively small evolution with redshift.
Comparing these super-sample errors to the BOSS DR12 measurement, we find that
the long modes have a much smaller effect $\sim0.1\%$ than the quoted precision
$\sim1.5\%$ on $\Dist{A}(z)$ in all redshift bins.
Regarding $H(z)$, the super-survey modes make a relatively bigger impact
$\sim0.3\%$, but is only of marginal significance compared to the BOSS
precision $\sim2\%$.
Similarly, the effect on $f\sigma_8$ at $\sim1\%$ is much smaller than the $\sim8\%$
precision of BOSS measurement.

Future surveys like DESI \cite{AghamousaEtAl16a} will put tighter constraint on
parameters.
While the super-sample effect is likely to remain negligible for the BAO, it
can become important for RSD measurements.
Comparing to the three BOSS redshift bins \cite{AlamEtAl17}, the DESI redshift
bins between $0.6$ and $1.0$ of width $\delta z=0.1$ have comparable volume and
therefore should have similar super-sample error on $f\sigma_8$.
However, its forecasted error is $2\%\sim3.3\%$ with $k_\mathrm{max}=0.1\hMpc$,
and $1\%\sim1.6\%$ with $k_\mathrm{max}=0.2\hMpc$.
The super-sample error becomes comparable in size in the latter case,
whose forecast seems optimistic for now but promising with recent
improvements of nonlinear reconstruction techniques \cite{ZhuYuEtAl16,
SchmittfullEtAl17, ShiCautunLi17} and forward modelling
\cite{SeljakEtAl17}.
Therefore we expect the super-sample effect may be non-negligible for RSD
measurements from future surveys such as DESI.

\section{Conclusion}
\label{sec:con}

We consider the impact of long-wavelength modes on the clustering of galaxies
in redshift space.
It has been well known that the mean density fluctuation in a finite volume
will change the local growth and expansion, named the growth effect and the
dilation effect respectively.
In this paper we have shown that in redshift space both the mean density and tidal modes modulate the local
growth and expansion in an angle-dependent way.
Out of 5 independent tidal components, only the one along the LOS remains after
azimuthal average in redshift space, thus one only needs to consider two
long-wavelength modes: the mean density $\Delta_0$ and LOS-projection of the
mean tide $\Delta_2$.
We have computed the response of the redshift-space galaxy power spectrum
$P(k,\mu)$ to these long modes, at leading order in standard perturbation
theory.
Our results are in \eqref{Rmu_sup}, \eqref{GD_mu}, \eqref{GD_mu2}, and alternatively
when written as responses of multipoles $P_l(k)$, in \eqref{Rl_sup},
Tabs.~\ref{tab:G_l}, \ref{tab:D_l}, and Fig.~\ref{fig:Rl_ana}.
The dilation effect alone coherently shifts all scales according to
\eqref{dilation}, and thus will affect the BAO measurement and the AP effect.
This cannot be removed with reconstruction techniques because the effect
originates from long modes outside the survey.

In order to verify our analytic results on linear scales, as well as to
measure the responses in the quasi-linear regime, we have performed a numerical
analysis on about a thousand QPM mocks.
Using sub-volumes of two different sizes and shapes, we have estimated both the
long modes and the power spectra of those regions.
By correlating the two we obtain the nonlinear galaxy response functions.
Our numerical (Figs.~\ref{fig:Rl_glb} and \ref{fig:Rl_loc}) and analytic
results agree well with each other on large scales, and differ on smaller
scales due to the nonlinear RSD.

The galaxy response functions derived in this paper can find applications in
the tidal reconstruction method \cite{PenShethEtAl12,ZhuPenEtAl16} to measure
the large-scale tidal field.
For optimal extraction of information it is critical to accurately calibrate
the response functions into the nonlinear regime using simulations.
Separate universe simulations \cite{LiHuEtAl14, McDonald03, Sirko05,
GnedinKravtsovEtAl11, BaldaufSeljakEtAl11a, WagnerSchmidtEtAl15, LiHuEtAl16,
LazeyrasWagnerEtAl16, BaldaufSeljakEtAl16, HuEtAl16, ChiangEtAl16,
ChiangEtAl17, CieplakSlosar16} provide an efficient method to measure the response of various
observables to the long density mode, such as power spectrum and halo mass
function, without sample variance or shot noise.
In a similar fashion, the mean tidal mode can be absorbed in the local
simulation background at the Newtonian level.
We leave this work on tidal simulations and reconstruction for future studies.

The long modes beyond the scale of a survey coherently modulate
the galaxy clustering according to their response functions \eqref{Cll_SS},
leading to an excess power spectrum covariance known as the super-sample
covariance.
This increased error loosens the constraints on cosmological parameters.
Compared to other covariance components, the SSC term scales
roughly at the same rate with the volume and thus its effect remains even for
much bigger surveys.
Because the sub-volume of QPM mocks have non-vanishing long modes, whereas the full
periodic volume does not, we have measured the SSC term by comparing their
covariance matrices, and confirmed its consistency with the numerical response
functions (see Fig.~\ref{fig:Cl_Kdiag} and \ref{fig:Cl_Koffd}).
Interestingly, while the SSC contributes significantly to the monopole $P_0$
auto-covariance, it seems negligible in the other covariance blocks, hinting at
possible strong angular dependence of the redshift-space galaxy trispectrum.

To assess the impact of the super-sample effect on galaxy clustering,
we adopt a simple analytic approach that translates the amplitudes of the long
modes to their induced systematic error on observables when ignoring these long modes.
To find their typical amplitudes, we have estimated the variance of the
super-survey long modes and their correlation (Tab.~\ref{tab:sigmaLL_boss})
given the survey window represented by a random catalog.
Taking BOSS DR12 as an example, we have found that neglecting the super-sample
effect can cause sub-percent to percent levels of systematic error on various
parameters constrained by BAO, AP, and RSD measurements, as listed in
Tab.~\ref{tab:impact}.
Compared to the BOSS DR12 error bars, which are at percent to ten percent levels, this
implies that the super-sample effect has only a marginal impact on the constraining
power of current spectroscopic galaxy surveys.
However our results suggest that for DESI
the super-sample error on $f\sigma_8$ may become comparable in size to
the forecasted error, especially when more small-scale information is made
available by iterative reconstruction and forward modelling methods.
We therefore expect that the super-sample effect can be non-negligible for RSD measurements
in future surveys.

\acknowledgments{YL thanks Jeremy Tinker and Martin White for help with the QPM
mocks.
YL acknowledges support from Fellowships at the Berkeley Center for
Cosmological Physics, and at the Kavli IPMU established by World Premier
International Research Center Initiative (WPI) of the MEXT, Japan.
MS acknowledges support from the Bezos Fund.
We acknowledge support from NASA grant NNX15AL17G.

This research used resources of the National Energy Research Scientific
Computing Center, a DOE Office of Science User Facility supported by the Office
of Science of the U.S.~Department of Energy under Contract No.~DE-AC02-05CH11231.
}

\appendix

\section{Standard perturbation theory}
\label{sec:spt}

In standard perturbation theory \cite{BernardeauColombiEtAl02}, the tree-level
matter bispectrum is determined by the $F_2$ kernel and the linear matter power
spectrum $\Plin$
\begin{equation}
    \label{Btree_real}
    B(\vk_1, \vk_2, \vk_3) = 2 F_2(\vk_1, \vk_2) \Plin(k_1) \Plin(k_2)
    + 2\,\mathrm{perm}.
\end{equation}
The $F_2$ kernel is
\begin{equation}
    \label{F2}
    F_2(\vk,\vk')
        = \frac{17}{21}
        + \frac{\vk\cdot\vk'}2 \Bigl(\frac1{k^2}+\frac1{k'^2}\Bigr)
        + \frac4{21} \L_2(\uvk\cdot\uvk'),
\end{equation}
with the 3 pieces named growth, shift, and tidal terms respectively.

For a galaxy field in redshift space we can introduce the second order biasing
and linear RSD effects by using the $Z_1$ and $Z_2$ kernels
\cite{ScoccimarroCouchmanEtAl99, Gil-MarinWagnerEtAl14}.
And the tree-level bispectrum of one matter and two galaxy fields is
\begin{multline}
    \label{Btree_red}
    B_{\matter\gal\gal}(\vk_1, \vk_2, \vk_3)
    = 2 Z_2(\vk_1, \vk_2) Z_1(\vk_2) \Plin(k_1) \Plin(k_2)
    + 2 Z_2(\vk_1, \vk_3) Z_1(\vk_3) \Plin(k_1) \Plin(k_3) \\
    + 2 F_2(\vk_2, \vk_3) Z_1(\vk_2) Z_1(\vk_3) \Plin(k_2) \Plin(k_3).
\end{multline}
The redshift-space kernels are
\begin{align}
    \label{Z1}
    Z_1(\vk) &= b_1+f\frac{(\vk\cdot\los)^2}{k^2} \\
    Z_2(\vk,\vk') &= b_1F_2(\vk,\vk')
        + \frac{b_1f(\vk+\vk')\cdot\los}2
        \Bigl(\frac{\vk\cdot\los}{k^2}+\frac{\vk'\cdot\los}{k'^2}\Bigr)
        \nonumber\\
    &\quad + \frac{f^2}2 \frac{\bigl[(\vk+\vk')\cdot\los\bigr]^2}{kk'}
        \frac{(\vk\cdot\los)(\vk'\cdot\los)}{kk'}
        + f\frac{\bigl[(\vk+\vk')\cdot\los\bigr]^2}{(\vk+\vk')^2}
            G_2(\vk,\vk') \nonumber\\
    &\quad + \frac{b_2}2 + \frac{b_{s^2}}2S_2(\vk,\vk')
    \label{Z2}
\end{align}
where $\los$ is the LOS direction and
\begin{align}
    \label{G2}
    G_2(\vk,\vk')
        &= \frac{13}{21}
        + \frac{\vk\cdot\vk'}2 \Bigl(\frac1{k^2}+\frac1{k'^2}\Bigr)
        + \frac8{21} \L_2(\uvk\cdot\uvk') \\
    S_2(\vk,\vk') &= \Bigl(\frac{\vk\cdot\vk'}{kk'}\Bigr)^2 - \frac13
    = \frac23 \L_2(\uvk\cdot\uvk')
    \label{S2}
\end{align}
Note that the forms of $F_2$ and $G_2$ kernels are only exact for flat
matter-dominated (Einstein-de Sitter) universe, but are good
approximations for cosmologies with $\Omega_\matter/f^2\approx1$.

To evaluate the real-space squeezed 3-point correlation in \eqref{3pt_real} at
tree level, we need to expand the $F_2$ kernel of the following forms in $q\ll
k$ limit
\begin{equation}
    \label{F2_exp}
    F_2(\vk+\vq_1,\vq_2) = \frac{17}{21} + \frac{(\vk+\vq_1)\cdot\vq_2}{2q_2^2}
    + \frac4{21} \L_2(\uvk\cdot\uvq_2) + \order(q/k).
\end{equation}
Because $F_2$ starts with $\order(k/q)$, we should expand the linear power
spectrum to $\order(q/k)$
\begin{equation}
    \label{Plin_exp}
    \Plin(\vk+\vq) = (1+\vq\cdot\nabla_\vk) \Plin(k) + \order(q^2/k^2)
    = \Bigl(1+\frac{\vq\cdot\vk}{k^2} \frac{\d\ln\Plin}{\d\ln k}\Bigr) \Plin(k)
    + \order(q^2/k^2),
\end{equation}
so that their combinations expands to $\order(1) \times \Plin(k)\Plin(q)$,
accounting for 2 terms of the 3 permutations in \eqref{Btree_real}.
The remaining piece takes roughly the form $F_2(\vk, -\vk-\vq) \Plin(k)
\Plin(|\vk+\vq|)$, thus is approximately $\order(q^2/k^2) \times \Plin^2(k)$
when $q\ll k$.
Therefore it is much smaller than the other 2 terms given that $P(k)/k^2$ is
monotonically decreasing, and we've dropped it in deriving \eqref{Btree_real1}.

Similarly for calculation in redshift space, we use the following kernel
expansions
\begin{align}
    \label{Z1_exp}
    Z_1(\vk+\vq) &= b_1+f\frac{(\vk\cdot\los)^2}{k^2}
        \Bigl[1+2\frac{\vq\cdot\los}{\vk\cdot\los}-2\frac{\vq\cdot\vk}{k^2}\Bigr]
        + \order(q^2/k^2), \\
    Z_2(\vk+\vq_1,\vq_2)
        %
        %
        &= b_1F_2(\vk+\vq_1,\vq_2)
        + \frac{b_1f}2 \biggl\{ \frac{(\vk\cdot\los)^2}{k^2}
            + \frac{\bigl[(\vk+\vq_1+\vq_2)\cdot\los\bigr](\vq_2\cdot\los)}{q_2^2} \biggr\} \nonumber\\
        &\quad + \frac{f^2}2 \frac{(\vk\cdot\los)^3(\vq_2\cdot\los)}{k^2q_2^2}
            \biggl[1+\frac{(3\vq_1+2\vq_2)\cdot\los}{\vk\cdot\los}-2\frac{\vq_1\cdot\vk}{k^2}\biggr] \nonumber\\
        &\quad + f\frac{(\vk\cdot\los)^2}{k^2} G_2(\vk+\vq_1,\vq_2)
            + f\frac{(\vk\cdot\los)^2}{k^2} \frac{\vk\cdot\vq_2}{q_2^2}
            \biggl[\frac{(\vq_1+\vq_2)\cdot\los}{\vk\cdot\los}-\frac{(\vq_1+\vq_2)\cdot\vk}{k^2}\biggr] \nonumber\\
        &\quad + \frac{b_2}2 + \frac{b_{s^2}}2S_2(\vk+\vq_1,\vq_2) + \order(q/k),
    \label{Z2_exp}
\end{align}
where
\begin{align}
    G_2(\vk+\vq_1,\vq_2) &= \frac{13}{21} + \frac{(\vk+\vq_1)\cdot\vq_2}{2q_2^2}
    + \frac8{21} \L_2(\uvk\cdot\uvq_2) + \order(q/k), \\
    S_2(\vk+\vq_1,\vq_2) &= S_2(\vk, \vq_2) + \order(q/k).
\end{align}
The $Z_1$ and $\Plin$ factors always occur together in \eqref{Btree_red}, so
it's useful to expand this combination using \eqref{Plin_exp} and
\eqref{Z1_exp}
\begin{multline}
    \label{Z1Plin_exp}
    Z_1(\vk+\vq)P(\vk+\vq) = \biggl\{
        \Bigl[b_1+f\frac{(\vk\cdot\los)^2}{k^2}\Bigr]
            \Bigl[1+\frac{\vq\cdot\vk}{k^2} \frac{\d\ln P}{\d\ln k}\Bigr] \\
        + 2f\frac{(\vq\cdot\los)(\vk\cdot\los)}{k^2}
        -2f\frac{(\vk\cdot\los)^2}{k^2}\frac{\vq\cdot\vk}{k^2}
    \biggr\} P(k)
    + \order(q^2/k^2)
\end{multline}

\section{Estimating covariance of long modes}
\label{sec:long_cov}

\begin{figure}[htpb]
    \centering
    \includegraphics[width=3.6in]{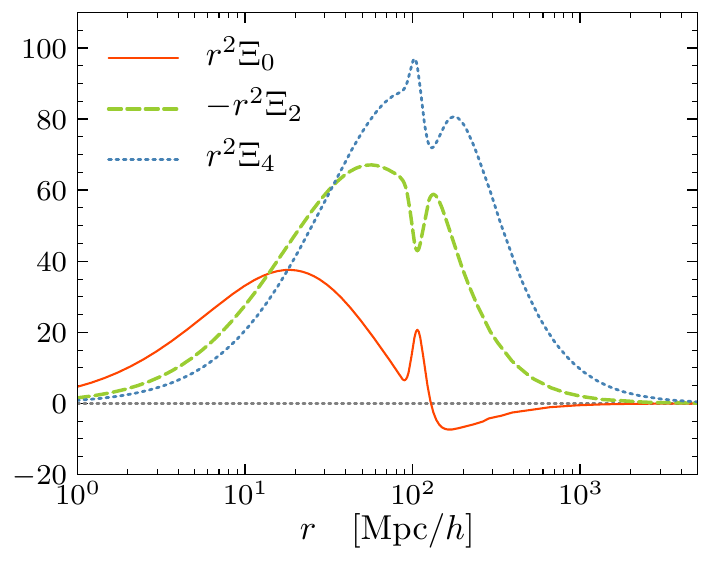}
    \caption{$\Xi_L(r)$ defined in \eqref{XiL}.
    We evaluate them with the Python package \texttt{mcfit} that implements the
    FFTLog algorithm.}
    \label{fig:XiL}
\end{figure}

Now let's compute the covariance of $\Delta_L$ for $L=0,2$ and some window
function $W$, given by \eqref{long_cov}.
Spell out each element and expand $\L_2^2$ in $\sigma_{22}$ in Legendre series
\begin{align}
    \sigma_{00}
    &= \frac{1}{V_2^2} \int_\vq P(q) \bigl|\WL(\vq)\bigr|^2, \nonumber\\
    \label{long_cov_1}
    \sigma_{02} = \sigma_{20}
    &= \frac{1}{V_2^2} \int_\vq P(q) \bigl|\WL(\vq)\bigr|^2
    \L_2(\uvq\cdot\los), \\\nonumber
    \sigma_{22}
    &= \frac{1}{V_2^2} \int_\vq P(q) \bigl|\WL(\vq)\bigr|^2
    \Bigl[\frac{18}{35}\L_4(\uvq\cdot\los) + \frac27\L_2(\uvq\cdot\los) + \frac15\Bigr].
\end{align}
And the Pearson coefficient measures the linear correlation between $\Delta_0$
and $\Delta_2$
\begin{equation}
    \rho_{02} \equiv \frac{\sigma_{02}}{\sqrt{\sigma_{00}\sigma_{22}}}.
\end{equation}

In order to compute \eqref{long_cov_1}, one needs to evaluate Fourier space
integrals of the following form in \eqref{long_cov_int}.
However this is not easy in Fourier space, as one needs a densely sampled large
FFT grid to well resolve and accurately measure $\WL(\vq)$.
So to perform the numerical integration we can rewrite it as a double integral
in configuration space
\begin{align}
    \frac{1}{V_2^2} \int_\vq P(q) \bigl|\WL(\vq)\bigr|^2 \L_L(\uvq\cdot\los)
    &= \frac{1}{V_2^2} \int\!\frac{q^2\d q}{2\pi^2}\int_\uvq P(q) \L_L(\uvq\cdot\los)
    \int_{\vx\vr} \WL(\vx)\WL(\vx+\vr) e^{-i\vq\cdot\vr} \nonumber\\
    &= \frac{1}{V_2^2} \int_{\vx\vr} \Xi_L(r)
    \WL(\vx)\WL(\vx+\vr) \L_L(\uvr\cdot\los),
    \label{long_cov_int}
\end{align}
before discretizing it as a double sum.
Here we have defined
\begin{equation}
    \label{XiL}
    \Xi_L(r) \equiv (-i)^L \int\!\frac{q^2\d q}{2\pi^2}\, j_L(qr) P(q),
\end{equation}
where $j_L$ is the spherical Bessel function of order $L$.
Note that $\Xi_0$ is just the correlation function $\xi$, but for higher order
$\Xi_L$ is different from the correlation function multipoles $\xi_L$.
$\Xi_L(r)$ can be evaluated most efficiently and accurately with the FFTLog
algorithm \cite{Hamilton00}.
For this we use the publicly available Python package
\texttt{mcfit}\footnote{\url{http://github.com/eelregit/mcfit}} and show the
result in Fig.~\ref{fig:XiL}.

In practice window functions are conveniently approximated by random catalogs
of points $\vx_i$ for $i=1,\cdots,N_\mathrm{r}$, whose distribution follows the
mean number density of galaxies $\nbar(\vx)$.
Thus we can exploit them to discretize the configuration space integrals into
sums over random points, by replacing
$\int_\vx\nbar\to\frac{N_\gal}{N_\ran}\sum_{i=1}^{N_\ran}$.
Recall that $\WL=W^2=\nbar^2w^2$.
Therefore we can Monte-Carlo the integral \eqref{long_cov_int}
\begin{align}
    \frac{1}{V_2^2} \int_\vq P(q) \bigl|\WL(\vq)\bigr|^2 \L_L(\uvq\cdot\los)
    &\approx \frac{\sum_{i\neq j} \Xi_L(r)
        \nbar(\vx_i)\nbar(\vx_j)w^2(\vx_i)w^2(\vx_j) \L_L(\uvr\cdot\los)
    }{\bigl[\sum_i \nbar(\vx_i)w^2(\vx_i)\bigr]^2}.
\end{align}
where we have removed the shot noise in $|\WL(\vq)|^2$ by omitting the $i=j$
terms in the numerator.
To test our method, we populate an $8\Mpch$ tophat with uniform random sample
to compute $\sigma_8$, and find that only $1000$ random points are needed to
reach percent level accuracy.

\bibliographystyle{JHEP}
\bibliography{cosmo,cosmo_preprints}
\end{document}